\documentclass[journal,11pt,draftclsnofoot,onecolumn]{IEEEtran}
\IEEEoverridecommandlockouts

\usepackage{amsmath,amssymb,amsfonts}
\usepackage{algorithm}
\usepackage{algorithmic}
\usepackage{graphicx}
\usepackage{booktabs}
\usepackage{enumitem}
\usepackage{url}
\usepackage{xcolor}
\usepackage{hyperref}
\usepackage{float}

\graphicspath{{fig/}}

\newcommand{\R}{\mathbb{R}}
\newcommand{\T}{\mathrm{T}}

\newcommand{\wrap}{\operatorname{wrap}}
\newcommand{\clip}{\operatorname{clip}}
\newcommand{\diag}{\operatorname{diag}}
\newcommand{\codeurl}{\url{https://github.com/ShineMinxing/PythonYOLO}}

\begin{document}

\title{Image-Domain Tilt Constrained Distributed Fusion for Maneuvering UAV Tracking with Multi-Camera Electro-Optical Observations}

\author{
Minxing~Sun and Yao~Mao
\thanks{Minxing Sun and Yao Mao are with the Institute of Optics and Electronics, Chinese Academy of Sciences, Chengdu, China.}
\thanks{Minxing Sun is also with the Institute for Infocomm Research (I\textsuperscript{2}R), Agency for Science, Technology and Research, Singapore, and Shenzhen Astralldynamics Technology.}
\thanks{Corresponding email: {\tt\small sunminxing20@mails.ucas.ac.cn, maoyao@ioe.ac.cn}.}
}

\maketitle

\begin{abstract}
Short-horizon prediction is essential for electro-optical UAV tracking, especially when the target is small, maneuvering, or intermittently observed.
Image center, line-of-sight, and range measurements provide direct constraints on target position, but their constraints on acceleration are weak.
As a result, prediction can lag during aggressive maneuvers.

This paper proposes an image-domain tilt constrained distributed fusion method for maneuvering UAV tracking.
The method uses the apparent roll and pitch of a rotorcraft target in the image as low-level maneuver cues.
A weak-prior auto-labeling pipeline first generates oriented bounding box and image-domain tilt labels from synchronized video, gimbal IMU, and UAV IMU data.
A YOLO-OBB detector is then trained to provide online target position and tilt measurements.
The front-end Python implementation is publicly available at \codeurl.

In the fusion stage, the UAV state is modeled by position, velocity, and acceleration.
Image-domain roll and pitch are introduced as acceleration-related pseudo-observations.
For distributed tracking, one mobile gimbal camera and two fixed ground cameras are fused asynchronously.
Camera attitude error states are augmented into the filter to absorb extrinsic drift and cross-camera systematic inconsistency.
A Mahalanobis gate with time-since-last-valid covariance widening is used to reject false detections and handle dropouts.

In simulation, adding roll/pitch observations reduces the prediction RMSE from 1.991 m to 0.821 m and decreases the cumulative prediction error by 60.75\%.
In real distributed experiments, a self-consistency evaluation shows an 18.10\% reduction in cumulative prediction error.
The results show that image-domain tilt can provide useful acceleration constraints for robust short-horizon UAV prediction.
\end{abstract}

\begin{IEEEkeywords}
UAV tracking, electro-optical tracking, image-domain tilt, YOLO-OBB, distributed fusion, multi-camera tracking, unscented Kalman filter
\end{IEEEkeywords}

\section{Introduction}

Small UAVs have been widely used in inspection, emergency response, public safety, and low-altitude operation scenarios.
Their increasing density and task diversity also bring new requirements for low-altitude monitoring, key-area protection, and safety management.
In these applications, a tracking system is expected not only to detect the target, but also to continuously output stable target states for pointing control, cooperative observation, and short-horizon prediction.

Electro-optical tracking is an important part of low-altitude monitoring systems because optical sensors provide high angular resolution and rich appearance information.
In practical systems, optical tracking is often combined with radar, radio-frequency, acoustic, or other wide-area sensors to form a coarse-to-fine monitoring chain \cite{DD2025}.
After the target enters the optical tracking stage, the key problem becomes whether the system can maintain a stable closed loop of image measurement, state estimation, prediction, and servo pointing.
This closed loop is affected by image noise, target scale variation, motion blur, background clutter, vibration, atmospheric disturbance, processing latency, and communication delay \cite{DB2019,AQ2025,SC2026,DM2025,RQ2025,OG2024}.
Therefore, the state estimator must provide a prediction that is aligned with the control or fusion time, rather than only smoothing past measurements.

UAV image detection has been extensively studied with deep neural networks.
Two-stage detectors such as Faster R-CNN provide strong region-level detection capability, while one-stage detectors represented by the YOLO family are more suitable for real-time deployment \cite{FR2017,DK2022,PJ2024}.
For UAV monitoring, bird-like distractors, small target size, low contrast, and changing backgrounds remain important causes of false detections and missed detections \cite{DK2022,PA2023,ST2025}.
Existing works improve detection robustness through transfer learning, lightweight network design, data augmentation, infrared information, and multi-dataset training \cite{LH2024,ED2025,ES2023,RS2025}.
However, a detector output becomes useful for electro-optical tracking only after it is converted into reliable measurements for state estimation and prediction.

Classical recursive estimators, including Kalman filtering and its nonlinear variants, provide a clear state-space framework for target tracking \cite{DT1999,LW1964,NK1960}.
For nonlinear optical measurements, extended Kalman filtering, unscented Kalman filtering, cubature Kalman filtering, and particle filtering have been widely investigated \cite{NK1960,EB2001,NJ1997,UW2000,UV2000,UJ2004,CA2009,NG1993,PF2001}.
For maneuvering targets, multiple-model methods such as IMM improve adaptability by maintaining several motion hypotheses \cite{IB1988,NF2010,ET2010}.
Robust and adaptive filters further address model mismatch, time-varying noise, and outliers by covariance adaptation, residual reweighting, or statistical gating \cite{RH1992,FZ1981,FS2001,RG2003,RJ2018,RW2015,CM2020,AT2024}.
Nevertheless, most optical tracking filters still rely mainly on position-like measurements, such as image center, line-of-sight direction, and range.
These measurements directly constrain target position, but only indirectly constrain velocity and acceleration.
When the target maneuvers quickly or short dropouts occur, the acceleration estimate can lag, and the prediction error can be amplified.

Rotorcraft UAVs provide an additional physical cue.
Their apparent roll and pitch are related to the thrust direction and therefore to the horizontal acceleration tendency.
Although image-domain tilt is not an exact body attitude measurement, it contains useful maneuver information.
This paper uses this cue as an acceleration-related pseudo-observation.
Different from methods that only improve the detector, the proposed method couples image-domain tilt recognition with a distributed state estimator.

Distributed observation is also important for low-altitude UAV tracking.
Multi-node systems can reduce blind areas and improve geometric observability, but they introduce asynchronous sampling, heterogeneous observation quality, communication delay, and cross-camera calibration inconsistency \cite{RL2022b}.
Mobile observation platforms further increase flexibility in complex terrain.
Legged or wheel-legged robots can carry optical sensors to viewpoints that are difficult for fixed stations or conventional wheeled vehicles to reach \cite{MB2018,MV2020,PG2022}.
In this work, one mobile gimbal camera and two fixed ground cameras are fused in a unified state-space model.
Camera attitude error states are introduced to compensate residual extrinsic inconsistency, and a per-camera Mahalanobis gate is used to handle false detections and dropouts.

The main contributions are:
\begin{itemize}[leftmargin=*, itemsep=0.25em, topsep=0.25em]
    \item \textbf{Weak-prior tilt labeling for UAV images.}
    A video--IMU auto-labeling pipeline is constructed to generate UAV oriented bounding boxes and image-domain roll/pitch labels without manual frame-by-frame OBB annotation.

    \item \textbf{Acceleration-constrained prediction filtering.}
    Image-domain roll and pitch are modeled as pseudo-observations related to target acceleration, improving acceleration observability during maneuvers and short missing intervals.

    \item \textbf{Asynchronous distributed multi-camera fusion.}
    One mobile gimbal camera and two fixed cameras are fused in a unified state-space model.
    Camera attitude error states are augmented to improve cross-camera consistency.

    \item \textbf{Robust update under dropouts and false detections.}
    A per-camera Mahalanobis gate with time-since-last-valid covariance widening is used to reject hard outliers and down-weight suspicious observations.

    \item \textbf{Public front-end implementation.}
    The Python workflow for motion point extraction, target tracing, weak label generation, and YOLO-OBB training is released at \codeurl.
\end{itemize}

The remainder of this paper is organized as follows.
Section~\ref{sec:system} defines the observation model.
Section~\ref{sec:labeling} presents weak-prior labeling and YOLO-OBB training.
Section~\ref{sec:fusion} introduces the tilt-constrained distributed fusion method.
Section~\ref{sec:experiments} reports simulation and real experiments.
Section~\ref{sec:conclusion} concludes the paper.

\section{System and Observation Model}
\label{sec:system}

\subsection{System Composition}

The system contains observation nodes, a UAV target, and a processing node.
During weak-prior label generation, the observation node is a gimbal camera with an IMU, and the target UAV also records IMU data.
During distributed fusion, the observation network contains three cameras:
Cam1 is a mobile gimbal camera mounted on a quadruped robot, and Cam2/Cam3 are fixed ground cameras.

Let the $k$-th image be $\mathbf{I}_k$.
The synchronized gimbal IMU and UAV IMU measurements are denoted by $\mathbf{m}^{G}_k$ and $\mathbf{m}^{U}_k$.
The weak-prior dataset is
\begin{equation}
    \mathcal{D}
    =
    \left\{
    \left(\mathbf{I}_k,\mathbf{m}^{G}_k,\mathbf{m}^{U}_k\right)
    \right\}_{k=1}^{N}.
\end{equation}

For online fusion, the mobile camera pose is computed from robot odometry, the static transform from the robot body to the gimbal, gimbal encoder readings, and the camera-gimbal transform.
The two fixed cameras use calibrated initial extrinsics.
Residual attitude errors are estimated online as augmented states.

\subsection{Camera Geometry}

We use a world frame $\{W\}$ and camera frames $\{C_i\}$, $i=1,2,3$.
For camera $i$, its world position is $\mathbf{c}^{(i)}_w$, and its attitude is ${}^{W}\mathbf{R}_{C_i}$.
For target position $\mathbf{p}_w=[X_w,Y_w,Z_w]^\T$, the target coordinate in the camera frame is
\begin{equation}
    \mathbf{p}^{(i)}_c
    =
    \left({}^{W}\mathbf{R}_{C_i}\right)^\T
    \left(
        \mathbf{p}_w-\mathbf{c}^{(i)}_w
    \right)
    =
    \begin{bmatrix}
        X^{(i)}_c & Y^{(i)}_c & Z^{(i)}_c
    \end{bmatrix}^{\T}.
    \label{eq:pc}
\end{equation}
The azimuth, elevation, and range are
\begin{equation}
\begin{aligned}
    \alpha^{(i)}
    &=
    \operatorname{atan2}\!\left(Y^{(i)}_c,X^{(i)}_c\right),\\
    \epsilon^{(i)}
    &=
    \operatorname{atan2}\!\left(
    Z^{(i)}_c,
    \sqrt{(X^{(i)}_c)^2+(Y^{(i)}_c)^2}
    \right),\\
    \rho^{(i)}
    &=
    \sqrt{(X^{(i)}_c)^2+(Y^{(i)}_c)^2+(Z^{(i)}_c)^2}.
\end{aligned}
\label{eq:az_el_range}
\end{equation}
In the MATLAB simulation, angular coordinates are mapped to image coordinates by the field-of-view model
\begin{equation}
    u^{(i)}
    =
    \frac{W_{\rm img}}{2}
    +
    \frac{\alpha^{(i)}}{\Theta_x^{(i)}}W_{\rm img},
    \qquad
    v^{(i)}
    =
    \frac{H_{\rm img}}{2}
    +
    \frac{\epsilon^{(i)}}{\Theta_y^{(i)}}H_{\rm img},
    \label{eq:uv_fov}
\end{equation}
where $W_{\rm img}=1920$, $H_{\rm img}=1080$, and $\Theta_x^{(i)},\Theta_y^{(i)}$ are the horizontal and vertical fields of view.

Each camera provides a five-dimensional observation block:
\begin{equation}
    \mathbf{z}^{(i)}_k=
    \begin{bmatrix}
        u^{(i)}_k &
        v^{(i)}_k &
        \rho^{(i)}_k &
        \eta^{(i)}_{{\rm roll},k} &
        \eta^{(i)}_{{\rm pitch},k}
    \end{bmatrix}^{\T}.
    \label{eq:z_block}
\end{equation}
The stacked three-camera observation is
\begin{equation}
    \mathbf{z}_k
    =
    \begin{bmatrix}
        \mathbf{z}^{(1)\T}_k &
        \mathbf{z}^{(2)\T}_k &
        \mathbf{z}^{(3)\T}_k
    \end{bmatrix}^{\T}
    \in\R^{15}.
\end{equation}

\subsection{Image-Domain Tilt}

Let the line-of-sight unit vector in the camera frame be
\begin{equation}
    \boldsymbol{\ell}^{(i)}_c
    =
    \frac{\mathbf{p}^{(i)}_c}
    {\|\mathbf{p}^{(i)}_c\|}.
\end{equation}
Equivalently, using the field-of-view mapping,
\begin{equation}
    \psi^{(i)}
    =
    \frac{u^{(i)}-W_{\rm img}/2}{W_{\rm img}}\Theta_x^{(i)},
    \qquad
    \vartheta^{(i)}
    =
    \frac{v^{(i)}-H_{\rm img}/2}{H_{\rm img}}\Theta_y^{(i)},
\end{equation}
\begin{equation}
    \boldsymbol{\ell}^{(i)}_c
    =
    \begin{bmatrix}
        \cos\vartheta^{(i)}\cos\psi^{(i)}\\
        \cos\vartheta^{(i)}\sin\psi^{(i)}\\
        \sin\vartheta^{(i)}
    \end{bmatrix}.
    \label{eq:los}
\end{equation}
The local image-plane right and up bases are
\begin{equation}
    \mathbf{e}^{(i)}_{r}
    =
    \frac{
    \begin{bmatrix}
        -\sin\psi^{(i)} & \cos\psi^{(i)} & 0
    \end{bmatrix}^{\T}
    }
    {
    \left\|
    \begin{bmatrix}
        -\sin\psi^{(i)} & \cos\psi^{(i)} & 0
    \end{bmatrix}^{\T}
    \right\|
    },
    \qquad
    \mathbf{e}^{(i)}_{u}
    =
    \frac{
        \mathbf{e}^{(i)}_{r}\times \boldsymbol{\ell}^{(i)}_c
    }
    {
        \left\|
        \mathbf{e}^{(i)}_{r}\times \boldsymbol{\ell}^{(i)}_c
        \right\|
    }.
\end{equation}
Given an equivalent UAV normal vector $\mathbf{n}^{(i)}_c$ in camera $i$, the image-domain roll and pitch are defined as
\begin{equation}
\begin{aligned}
    \eta^{(i)}_{\rm roll}
    &=
    \wrap
    \left[
    \operatorname{atan2}
    \left(
        \mathbf{n}^{(i)\T}_c\mathbf{e}^{(i)}_{r},
        \mathbf{n}^{(i)\T}_c\mathbf{e}^{(i)}_{u}
    \right)
    \right],
    \\
    \eta^{(i)}_{\rm pitch}
    &=
    -
    \arcsin
    \left(
        \clip
        \left(
            \mathbf{n}^{(i)\T}_c\boldsymbol{\ell}^{(i)}_c,\ -1,\ 1
        \right)
    \right).
\end{aligned}
\label{eq:img_tilt}
\end{equation}
These angles are treated as noisy image-domain maneuver cues rather than exact UAV body attitude.

\begin{figure}[H]
    \centering
    \begin{minipage}[b]{0.48\linewidth}
        \centering
        \includegraphics[width=\linewidth]{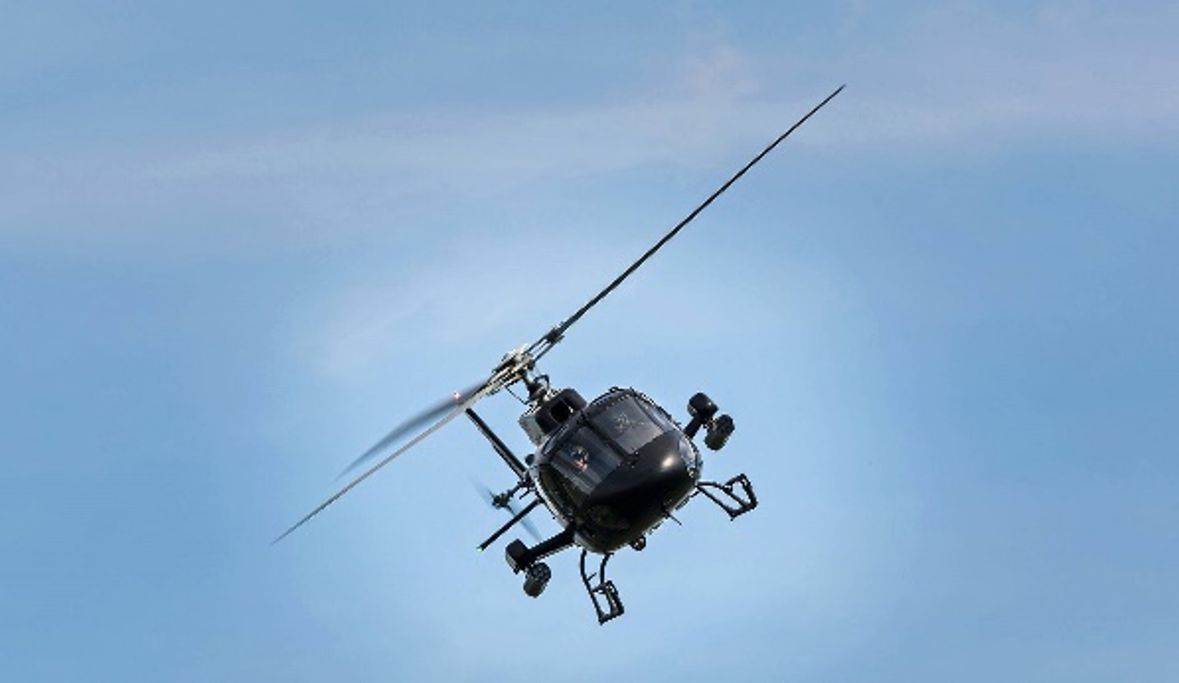}\\
        {\small (a) Helicopter tilt example.}
    \end{minipage}
    \hfill
    \begin{minipage}[b]{0.48\linewidth}
        \centering
        \includegraphics[width=\linewidth]{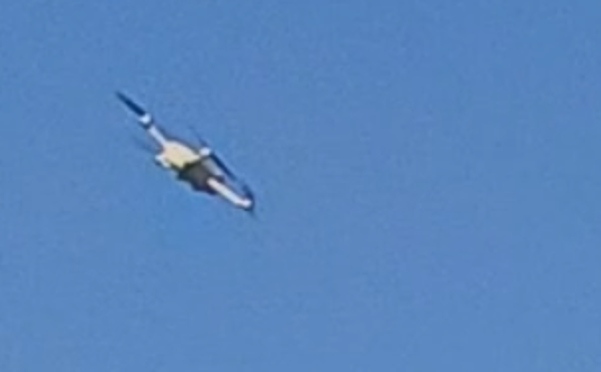}\\
        {\small (b) Quadrotor tilt example.}
    \end{minipage}
    \caption{Image-domain body tilt contains short-term maneuver tendency information.}
    \label{fig:motivation}
\end{figure}

\section{Weak-Prior Auto-Labeling and YOLO-OBB Tilt Recognition}
\label{sec:labeling}

\subsection{Overview}

The purpose of the front end is to obtain continuous target position and tilt measurements without manually labeling large numbers of OBB samples.
The workflow contains four steps:
motion candidate extraction, target trajectory association, weak roll/pitch label generation, and YOLO-OBB training.

\begin{figure}[htbp]
    \centering
    \begin{minipage}[b]{0.32\linewidth}
        \centering
        \includegraphics[width=\linewidth]{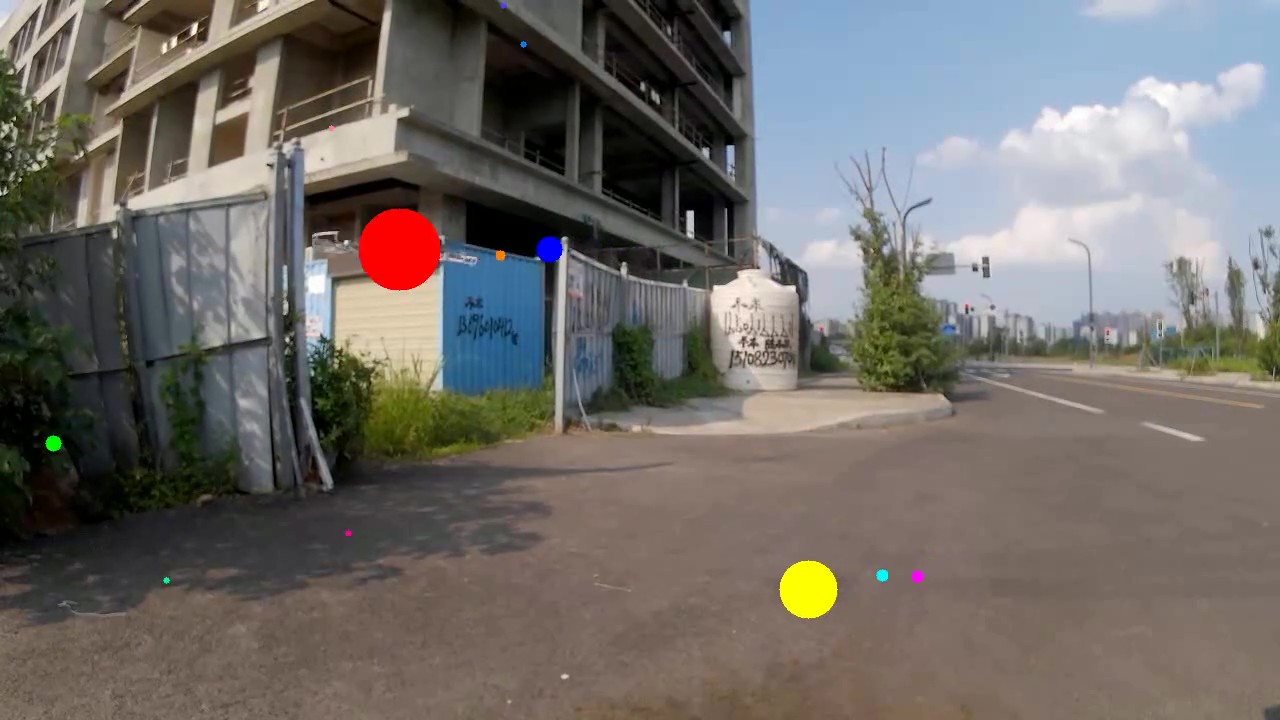}\\
        {\small (a) Motion candidates.}
    \end{minipage}
    \hfill
    \begin{minipage}[b]{0.32\linewidth}
        \centering
        \includegraphics[width=\linewidth]{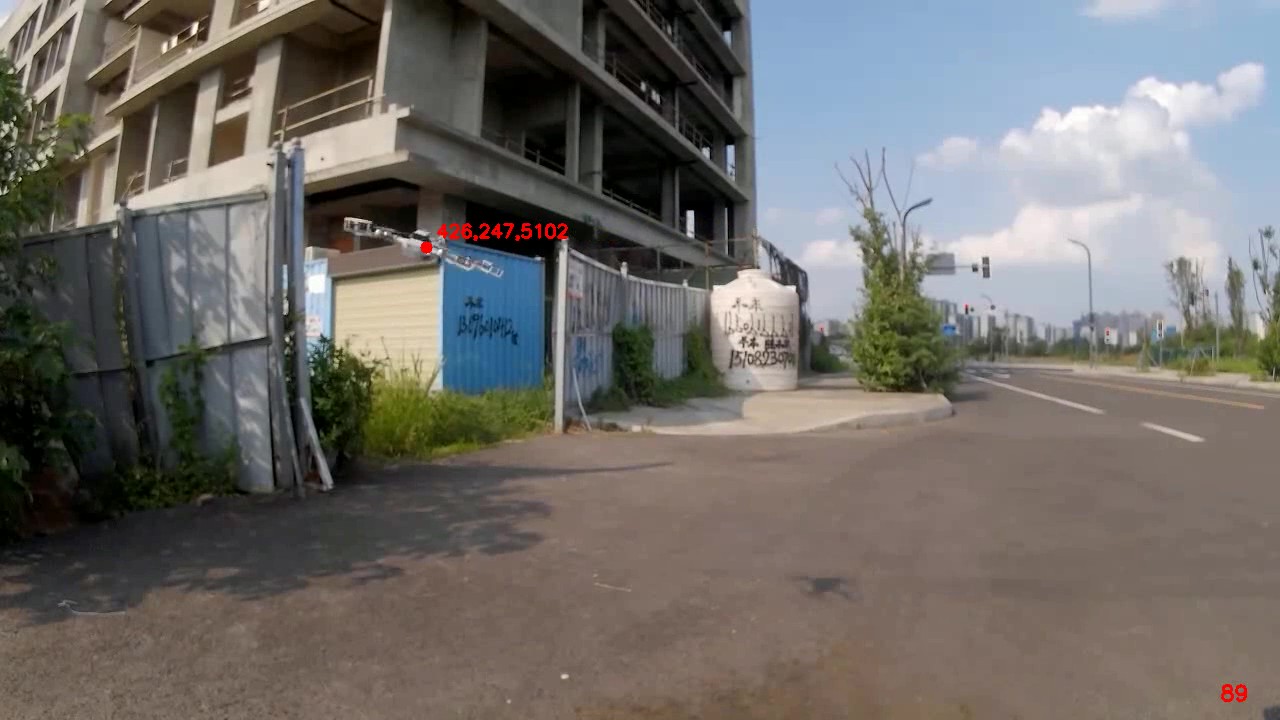}\\
        {\small (b) Tracked target.}
    \end{minipage}
    \hfill
    \begin{minipage}[b]{0.32\linewidth}
        \centering
        \includegraphics[width=\linewidth]{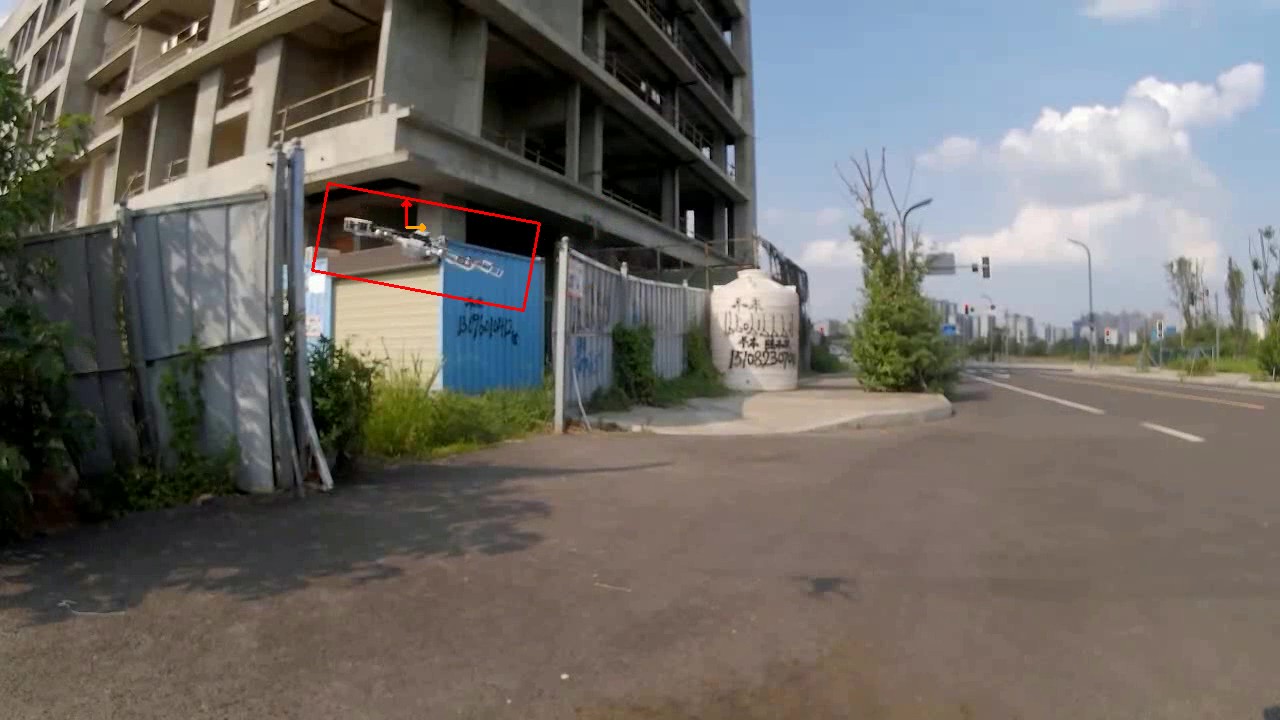}\\
        {\small (c) Weak tilt and OBB.}
    \end{minipage}
    \caption{Main steps of the weak-prior auto-labeling pipeline.}
    \label{fig:auto_label_pipeline}
\end{figure}

\subsection{Motion Candidate Extraction}

For each frame $\mathbf{I}_k$, a Gaussian mixture background model is used to obtain foreground candidates:
\begin{equation}
    p(\mathbf{p}_{{\rm pix},k})
    =
    \sum_{i=1}^{K}
    \omega_i
    \mathcal{N}
    \left(
        \mathbf{p}_{{\rm pix},k};
        \boldsymbol{\mu}_i,
        \boldsymbol{\Sigma}_i
    \right).
\end{equation}
A pixel is regarded as foreground if it cannot be explained by the current background model under a Mahalanobis threshold:
\begin{equation}
    \left(
    \mathbf{p}_{{\rm pix},k}-\boldsymbol{\mu}_i
    \right)^\T
    \boldsymbol{\Sigma}^{-1}_i
    \left(
    \mathbf{p}_{{\rm pix},k}-\boldsymbol{\mu}_i
    \right)
    >
    \tau_{\rm pix}.
\end{equation}
After morphology and connected-component extraction, a contour with area $A_{n,k}$ is retained if
\begin{equation}
    A_{\min}\le A_{n,k}\le \rho_{\max} W_{\rm img}H_{\rm img}.
\end{equation}
The retained candidate set is
\begin{equation}
    \mathcal{P}_k
    =
    \left\{
    (x_{i,k},y_{i,k},A_{i,k})
    \right\}_{i=1}^{N_k}.
\end{equation}

\subsection{Target Trajectory Association}

Let the previous target position be
$\hat{\mathbf{s}}_{k-1}=[\hat u_{k-1},\hat v_{k-1}]^\T$.
The distance from candidate $i$ to the previous target is
\begin{equation}
    \delta_{i,k}
    =
    \left\|
    \begin{bmatrix}
        x_{i,k} & y_{i,k}
    \end{bmatrix}^{\T}
    -
    \hat{\mathbf{s}}_{k-1}
    \right\|_2 .
\end{equation}
Within a search radius $r^{\rm trk}_k$, the target candidate is selected by
\begin{equation}
    i_k^{\star}
    =
    \arg\max_{i:\delta_{i,k}\le r^{\rm trk}_k}
    \left(A_{i,k},-\delta_{i,k}\right).
\end{equation}
If no candidate is found, the frame is marked as missing and the search radius is enlarged.
Short missing segments are filled by interpolation.

\subsection{Weak Tilt Labels from Dual IMU}

The UAV IMU provides the UAV attitude ${}^{W}\mathbf{R}_{U}(k)$.
The gimbal IMU and camera-gimbal calibration provide the camera attitude ${}^{W}\mathbf{R}_{C}(k)$.
For a constant UAV equivalent normal vector $\mathbf{n}^{U}_0$, its camera-frame expression is
\begin{equation}
    \mathbf{n}^{C}_k
    =
    \left({}^{W}\mathbf{R}_{C}(k)\right)^\T
    {}^{W}\mathbf{R}_{U}(k)
    \mathbf{n}^{U}_0 .
\end{equation}
Together with the image target position, Eq.~\eqref{eq:img_tilt} gives weak image-domain roll/pitch labels.
To remove a constant installation bias, the labels are expressed relative to the first valid frame:
\begin{equation}
\begin{aligned}
    \tilde{\eta}_{{\rm pitch},k}
    &=
    \eta_{{\rm pitch},k}
    -
    \eta_{{\rm pitch},1},\\
    \tilde{\eta}_{{\rm roll},k}
    &=
    \wrap
    \left(
    \eta_{{\rm roll},k}
    -
    \eta_{{\rm roll},1}
    \right).
\end{aligned}
\end{equation}

\begin{figure}[htbp]
    \centering
    \begin{minipage}[b]{0.48\linewidth}
        \centering
        \includegraphics[width=\linewidth]{fig_step3_drone_picture_RollPitch_0}\\
        {\small (a)}
    \end{minipage}
    \hfill
    \begin{minipage}[b]{0.48\linewidth}
        \centering
        \includegraphics[width=\linewidth]{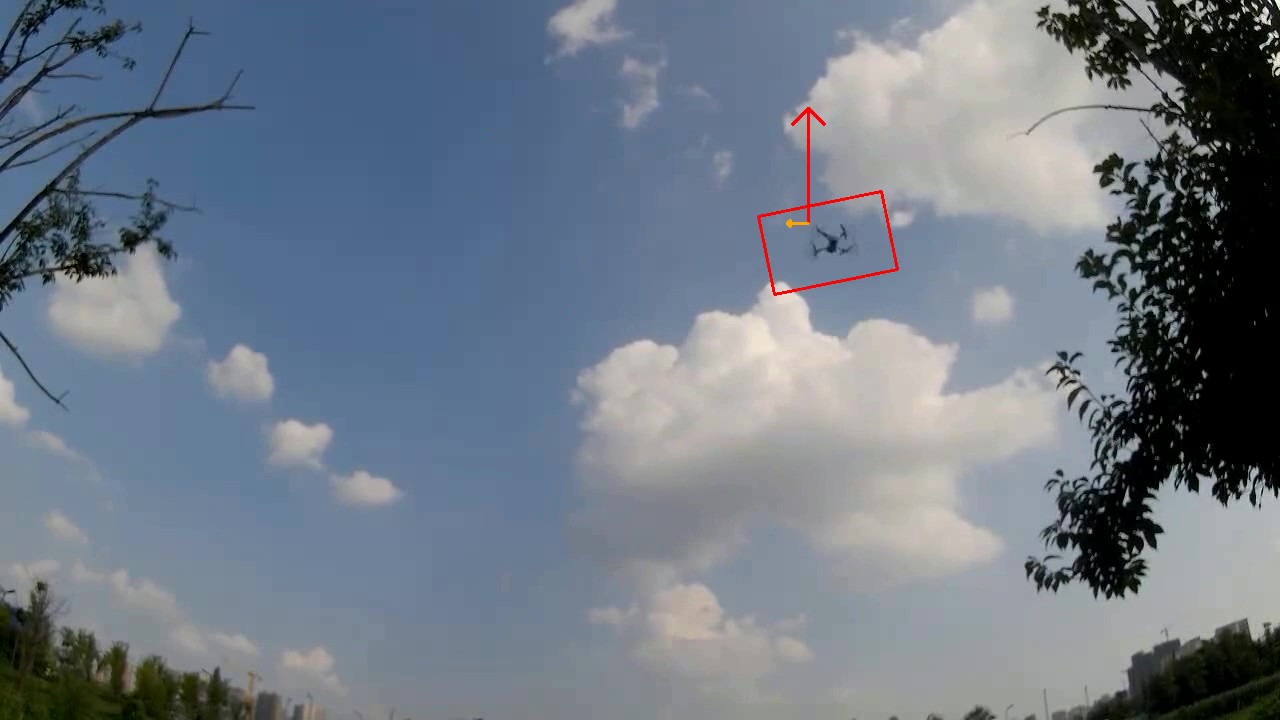}\\
        {\small (b)}
    \end{minipage}

    \vspace{0.4em}

    \begin{minipage}[b]{0.48\linewidth}
        \centering
        \includegraphics[width=\linewidth]{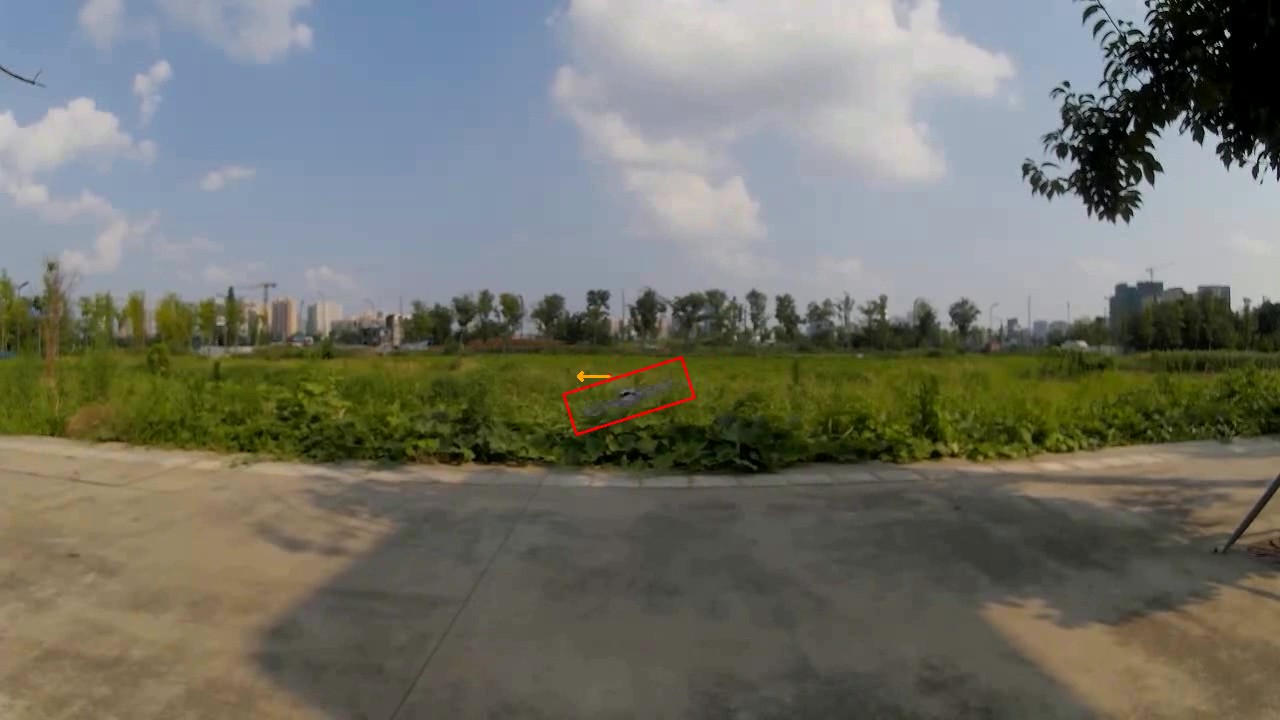}\\
        {\small (c)}
    \end{minipage}
    \hfill
    \begin{minipage}[b]{0.48\linewidth}
        \centering
        \includegraphics[width=\linewidth]{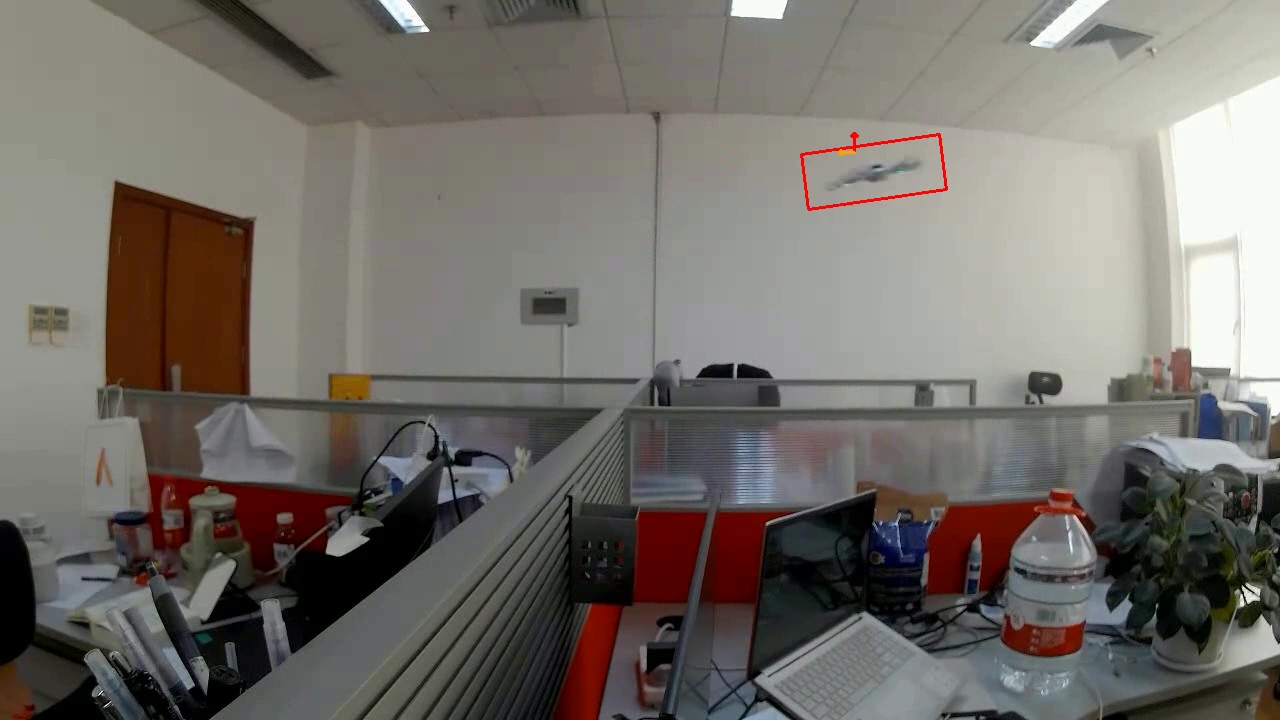}\\
        {\small (d)}
    \end{minipage}
    \caption{Examples of weak image-domain tilt labels and adaptive OBB labels.}
    \label{fig:weak_tilt_examples}
\end{figure}

\subsection{Adaptive OBB Label Generation}

Let the tracked target area be $A_{{\rm drone},k}$.
The OBB size is generated by
\begin{equation}
    W_{{\rm pix},k}
    =
    k_w\sqrt{A_{{\rm drone},k}},
    \qquad
    H_{{\rm pix},k}
    =
    k_h\sqrt{A_{{\rm drone},k}}
    \left(
    1+
    \left|
    \sin\tilde{\eta}_{{\rm pitch},k}
    \right|
    \right).
\end{equation}
Let $w_k=W_{{\rm pix},k}/2$ and $h_k=H_{{\rm pix},k}/2$.
The four vertices of the unrotated box are
\begin{equation}
    \mathbf{p}_{k,1}=
    \begin{bmatrix}w_k\\h_k\end{bmatrix},
    \quad
    \mathbf{p}_{k,2}=
    \begin{bmatrix}-w_k\\h_k\end{bmatrix},
    \quad
    \mathbf{p}_{k,3}=
    \begin{bmatrix}-w_k\\-h_k\end{bmatrix},
    \quad
    \mathbf{p}_{k,4}=
    \begin{bmatrix}w_k\\-h_k\end{bmatrix}.
\end{equation}
After rotation and translation,
\begin{equation}
    \bar{\mathbf{p}}_{k,j}
    =
    \mathbf{R}_{2}
    \left(
    -\tilde{\eta}_{{\rm roll},k}
    \right)
    \mathbf{p}_{k,j}
    +
    \begin{bmatrix}
        \hat u_k\\ \hat v_k
    \end{bmatrix},
\end{equation}
where
\begin{equation}
    \mathbf{R}_{2}(\alpha)=
    \begin{bmatrix}
        \cos\alpha & -\sin\alpha\\
        \sin\alpha & \cos\alpha
    \end{bmatrix}.
\end{equation}
The label is exported in four-point YOLO-OBB format.

\subsection{YOLO-OBB Training}

The generated dataset is used to train a YOLO-OBB detector.
The online detector outputs the UAV center, OBB geometry, and image-domain tilt cues.
The detector is trained not only for frame-level recognition, but also to provide stable low-level measurements for the subsequent state estimator.

\begin{algorithm}[htbp]
\caption{Weak-prior labeling and YOLO-OBB training}
\label{alg:auto_label}
\begin{algorithmic}[1]
\REQUIRE Video frames, gimbal IMU, UAV IMU, camera parameters
\ENSURE YOLO-OBB dataset and trained detector
\FOR{each frame $k$}
    \STATE Extract foreground candidates by background modeling.
    \STATE Filter contours by area and morphology.
\ENDFOR
\STATE Reconstruct the UAV image trajectory by radius-constrained association.
\STATE Interpolate short missing segments.
\FOR{each valid frame $k$}
    \STATE Compute the UAV normal in the camera frame from dual-IMU geometry.
    \STATE Generate weak image-domain roll/pitch labels.
    \STATE Generate an adaptive OBB from target area and weak tilt labels.
    \STATE Export the four-point YOLO-OBB annotation.
\ENDFOR
\STATE Train the YOLO-OBB detector.
\end{algorithmic}
\end{algorithm}

\section{Tilt-Constrained Distributed Fusion}
\label{sec:fusion}

\subsection{Augmented State}

The target motion state is represented by position, velocity, and acceleration:
\begin{equation}
    \mathbf{x}^{u}_k
    =
    \begin{bmatrix}
        \mathbf{p}^{\T}_k &
        \mathbf{v}^{\T}_k &
        \mathbf{a}^{\T}_k
    \end{bmatrix}^{\T}.
\end{equation}
The MATLAB fusion model uses an 18-dimensional augmented state:
\begin{equation}
\begin{aligned}
    \mathbf{x}_k
    =
    [
    &x_k,\dot{x}_k,\ddot{x}_k,
    y_k,\dot{y}_k,\ddot{y}_k,
    z_k,\dot{z}_k,\ddot{z}_k,\\
    &\phi^{(1)}_k,\theta^{(1)}_k,\psi^{(1)}_k,
    \phi^{(2)}_k,\theta^{(2)}_k,\psi^{(2)}_k,
    \phi^{(3)}_k,\theta^{(3)}_k,\psi^{(3)}_k
    ]^\T .
\end{aligned}
\label{eq:aug_state}
\end{equation}
The first nine states are the UAV motion states.
The last nine states are effective camera roll-pitch-yaw states.
They can be interpreted as nominal camera attitudes plus slowly varying error angles.

For each axis, the constant-acceleration propagation is
\begin{equation}
    \begin{bmatrix}
        q_{k+1}\\ \dot q_{k+1}\\ \ddot q_{k+1}
    \end{bmatrix}
    =
    \mathbf{F}_{\rm ca}(\Delta t_k)
    \begin{bmatrix}
        q_{k}\\ \dot q_{k}\\ \ddot q_{k}
    \end{bmatrix}
    +
    \mathbf{w}_{q,k},
\end{equation}
where
\begin{equation}
    \mathbf{F}_{\rm ca}(\Delta t_k)
    =
    \begin{bmatrix}
        1 & \Delta t_k & \frac{1}{2}\Delta t_k^2\\
        0 & 1          & \Delta t_k\\
        0 & 0          & 1
    \end{bmatrix}.
    \label{eq:f_ca}
\end{equation}
The camera attitude states use random-walk propagation with small process noise.

\subsection{Acceleration-to-Tilt Observation}

For a rotorcraft UAV, horizontal acceleration is related to the direction of the thrust vector.
A simplified relation is
\begin{equation}
    \eta_{\rm pitch}\approx \arctan\left(\frac{a_x}{g}\right),
    \qquad
    \eta_{\rm roll}\approx \arctan\left(\frac{a_y}{g}\right),
    \label{eq:acc_tilt_simple}
\end{equation}
where $g$ is the gravitational acceleration.
Instead of directly converting tilt to acceleration by a tangent inverse, the filter predicts the tilt from the acceleration states.
This avoids noise amplification at larger tilt angles.

In the implemented observation model, the acceleration states first define an equivalent UAV attitude:
\begin{equation}
    \theta_u = \arctan2(a_x, g),
    \qquad
    \phi_u = \arctan2(a_y, g),
    \qquad
    \psi_u = 0.
\end{equation}
The equivalent UAV normal vector in the world frame is
\begin{equation}
    \mathbf{n}_w
    =
    \mathbf{R}_z(\psi_u)
    \mathbf{R}_y(\theta_u)
    \mathbf{R}_x(\phi_u)
    \begin{bmatrix}
        0\\0\\-1
    \end{bmatrix}.
    \label{eq:normal_from_acc}
\end{equation}
For camera $i$,
\begin{equation}
    \mathbf{n}^{(i)}_c
    =
    \left({}^{W}\mathbf{R}_{C_i}\right)^\T
    \mathbf{n}_w.
\end{equation}
The predicted roll/pitch observation is then computed by Eq.~\eqref{eq:img_tilt}.

\subsection{Observation Prediction}

For each available camera, the state predicts the geometric observation
\begin{equation}
    \hat{\mathbf{z}}^{(i)}_{\rm geo}
    =
    \begin{bmatrix}
        \hat u^{(i)} &
        \hat v^{(i)} &
        \hat \rho^{(i)}
    \end{bmatrix}^{\T}
\end{equation}
by Eq.~\eqref{eq:pc}--\eqref{eq:uv_fov}.
The acceleration states predict the tilt observation
\begin{equation}
    \hat{\mathbf{z}}^{(i)}_{\eta}
    =
    \begin{bmatrix}
        \hat\eta^{(i)}_{\rm roll} &
        \hat\eta^{(i)}_{\rm pitch}
    \end{bmatrix}^{\T}.
\end{equation}
Thus,
\begin{equation}
    \hat{\mathbf{z}}^{(i)}
    =
    \begin{bmatrix}
        \hat u^{(i)} &
        \hat v^{(i)} &
        \hat \rho^{(i)} &
        \hat\eta^{(i)}_{\rm roll} &
        \hat\eta^{(i)}_{\rm pitch}
    \end{bmatrix}^{\T}.
\end{equation}

\subsection{Unscented Kalman Update}

The observation function contains camera projection, range geometry, angle wrapping, and acceleration-to-tilt mapping.
Therefore, a sigma-point nonlinear Kalman estimator is used.

Let the state dimension be $n$.
The sigma points are
\begin{equation}
\begin{aligned}
    \mathcal{X}_{0,k} &= \hat{\mathbf{x}}_{k|k},\\
    \mathcal{X}_{j,k} &= \hat{\mathbf{x}}_{k|k}
    +
    \left[
        \sqrt{(n+\lambda)\mathbf{P}_{k|k}}
    \right]_j,\quad j=1,\ldots,n,\\
    \mathcal{X}_{j+n,k} &= \hat{\mathbf{x}}_{k|k}
    -
    \left[
        \sqrt{(n+\lambda)\mathbf{P}_{k|k}}
    \right]_j,\quad j=1,\ldots,n .
\end{aligned}
\end{equation}
After propagation through the transition and observation functions,
\begin{equation}
    \hat{\mathbf{x}}_{k|k-1}
    =
    \sum_j W_j^{(m)}\mathcal{X}_{j,k|k-1},
\end{equation}
\begin{equation}
    \hat{\mathbf{z}}_{k|k-1}
    =
    \sum_j W_j^{(m)}\mathcal{Z}_{j,k}.
\end{equation}
The innovation covariance and cross covariance are
\begin{equation}
    \mathbf{P}_{zz}
    =
    \sum_j W_j^{(c)}
    \left(
        \mathcal{Z}_{j,k}-\hat{\mathbf{z}}_{k|k-1}
    \right)
    \left(
        \mathcal{Z}_{j,k}-\hat{\mathbf{z}}_{k|k-1}
    \right)^\T
    +
    \mathbf{R}_k,
\end{equation}
\begin{equation}
    \mathbf{P}_{xz}
    =
    \sum_j W_j^{(c)}
    \left(
        \mathcal{X}_{j,k|k-1}-\hat{\mathbf{x}}_{k|k-1}
    \right)
    \left(
        \mathcal{Z}_{j,k}-\hat{\mathbf{z}}_{k|k-1}
    \right)^\T .
\end{equation}
The update is
\begin{equation}
    \mathbf{K}_k
    =
    \mathbf{P}_{xz}\mathbf{P}_{zz}^{-1},
\end{equation}
\begin{equation}
    \hat{\mathbf{x}}_{k|k}
    =
    \hat{\mathbf{x}}_{k|k-1}
    +
    \mathbf{K}_k
    \left(
        \mathbf{z}_k-\hat{\mathbf{z}}_{k|k-1}
    \right),
\end{equation}
\begin{equation}
    \mathbf{P}_{k|k}
    =
    \mathbf{P}_{k|k-1}
    -
    \mathbf{K}_k\mathbf{P}_{xz}^{\T}.
\end{equation}
The updated state is then propagated to the prediction horizon $\tau$.

\subsection{Mahalanobis Gating}

Before the filter update, each camera block is gated independently.
The gate is applied to the geometric sub-block
\begin{equation}
    \mathbf{z}^{(i)}_{\rm geo}
    =
    \begin{bmatrix}
        u^{(i)} & v^{(i)} & \rho^{(i)}
    \end{bmatrix}^{\T}.
\end{equation}
Let the predicted geometric observation be $\hat{\mathbf{z}}^{(i)}_{\rm geo}$.
The residual is
\begin{equation}
    \boldsymbol{\nu}^{(i)}_k
    =
    \mathbf{z}^{(i)}_{{\rm geo},k}
    -
    \hat{\mathbf{z}}^{(i)}_{{\rm geo},k}.
\end{equation}
The effective covariance is
\begin{equation}
    \mathbf{R}^{(i)}_{\rm eff}
    =
    s_R\mathbf{R}^{(i)}_{0}
    +
    \mathbf{J}^{(i)}_p
    \mathbf{P}_{\rm pos}
    \mathbf{J}^{(i)\T}_p
    +
    \epsilon_R\mathbf{I},
    \label{eq:Reff}
\end{equation}
where $\mathbf{J}^{(i)}_p=\partial[u,v,\rho]/\partial[x,y,z]$ is computed by numerical difference.
The time-since-last-valid position uncertainty is
\begin{equation}
    \sigma_j
    =
    \kappa
    \left(
    |v_j|\Delta t
    +
    \frac{1}{2}|a_j|\Delta t^2
    \right),
    \qquad
    \mathbf{P}_{\rm pos}
    =
    \diag
    \left(
    \sigma_x^2,\sigma_y^2,\sigma_z^2
    \right).
    \label{eq:Ppos}
\end{equation}
The gating statistic is
\begin{equation}
    d_i^2
    =
    \boldsymbol{\nu}^{(i)\T}_k
    \left(
    \mathbf{R}^{(i)}_{\rm eff}
    \right)^{-1}
    \boldsymbol{\nu}^{(i)}_k .
    \label{eq:gating_stat}
\end{equation}
Two thresholds are used:
\begin{equation}
    \gamma_{\rm soft}=\chi^2_{3}(0.99),
    \qquad
    \gamma_{\rm hard}=\chi^2_{3}(0.999).
\end{equation}
The rule is
\begin{equation}
\begin{cases}
d_i^2>\gamma_{\rm hard}: & \text{reject camera block},\\
\gamma_{\rm soft}<d_i^2\le \gamma_{\rm hard}: & \text{accept with inflated covariance},\\
d_i^2\le \gamma_{\rm soft}: & \text{accept normally}.
\end{cases}
\end{equation}
If a camera block is accepted with inflated covariance, both its geometric and tilt covariances are inflated.
Thus, roll/pitch observations follow the same camera-level reliability decision.

\begin{algorithm}[htbp]
\caption{Tilt-constrained asynchronous multi-camera fusion}
\label{alg:fusion}
\begin{algorithmic}[1]
\REQUIRE Previous estimate, covariance, incoming camera observations
\ENSURE Updated state and short-horizon prediction
\STATE Assign observations to Cam1, Cam2, and Cam3 slots.
\STATE Propagate the augmented state by $\Delta t$.
\FOR{each camera slot $i$}
    \IF{camera block is missing}
        \STATE Mark camera $i$ unavailable.
    \ELSE
        \STATE Predict $[u,v,\rho,\eta_{\rm roll},\eta_{\rm pitch}]$.
        \STATE Compute the Mahalanobis statistic for $[u,v,\rho]$.
        \IF{$d_i^2>\gamma_{\rm hard}$}
            \STATE Reject this camera block.
        \ELSIF{$d_i^2>\gamma_{\rm soft}$}
            \STATE Accept with inflated covariance.
        \ELSE
            \STATE Accept normally.
        \ENDIF
    \ENDIF
\ENDFOR
\IF{at least one camera block is accepted}
    \STATE Perform the UKF measurement update.
\ENDIF
\STATE Propagate the updated state to the prediction horizon.
\end{algorithmic}
\end{algorithm}

\section{Experiments}
\label{sec:experiments}

\subsection{YOLO-OBB Training Results}

The generated dataset uses DOTA-style four-point OBB labels.
The UAV samples are generated by the weak-prior pipeline.
Additional categories and background samples are used to improve robustness in real scenes.

\begin{figure}[htbp]
    \centering
    \includegraphics[width=0.92\linewidth]{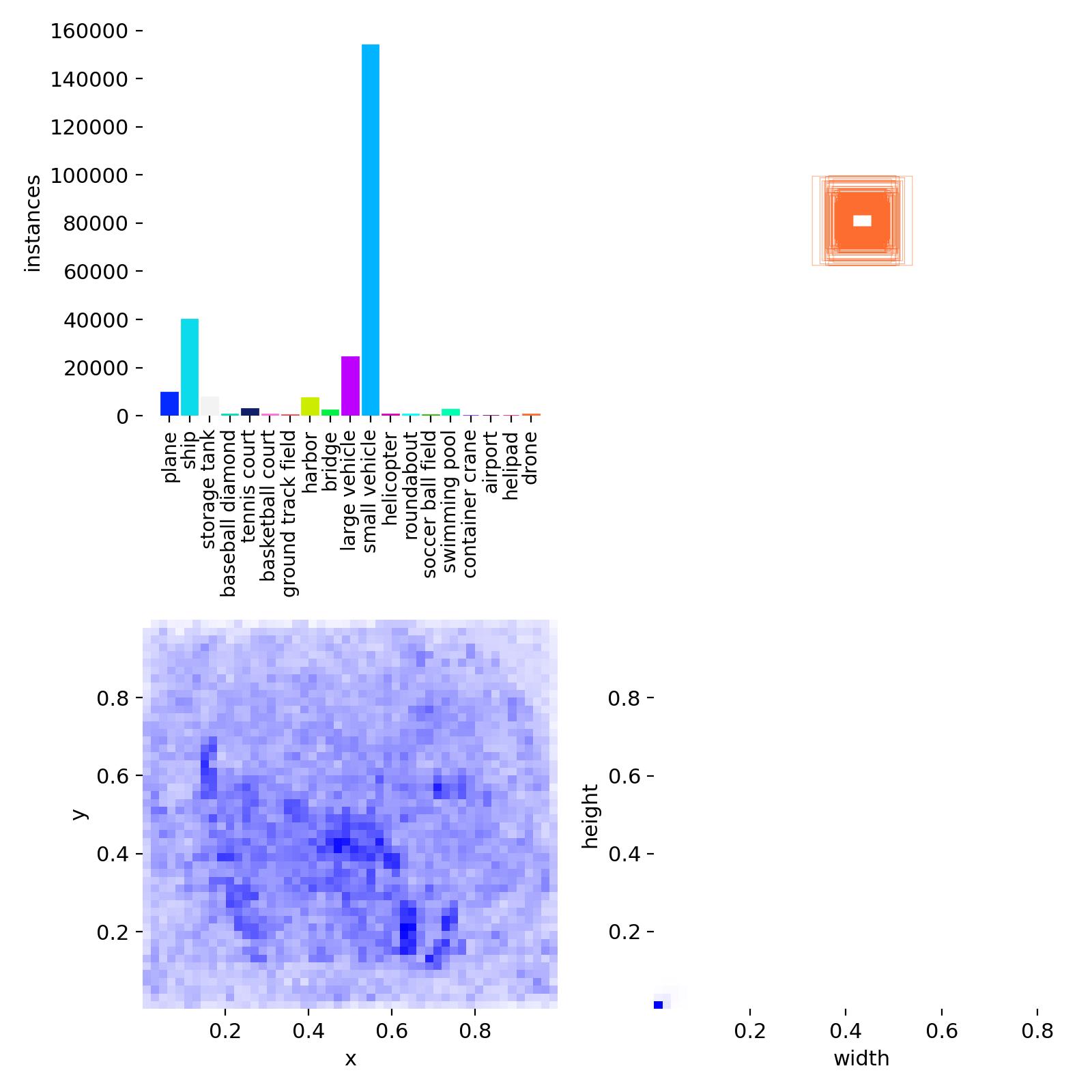}
    \caption{Class distribution and OBB label statistics of the generated dataset.}
    \label{fig:labels}
\end{figure}

\begin{figure}[htbp]
    \centering
    \begin{minipage}[t]{0.48\linewidth}
        \centering
        \includegraphics[width=\linewidth]{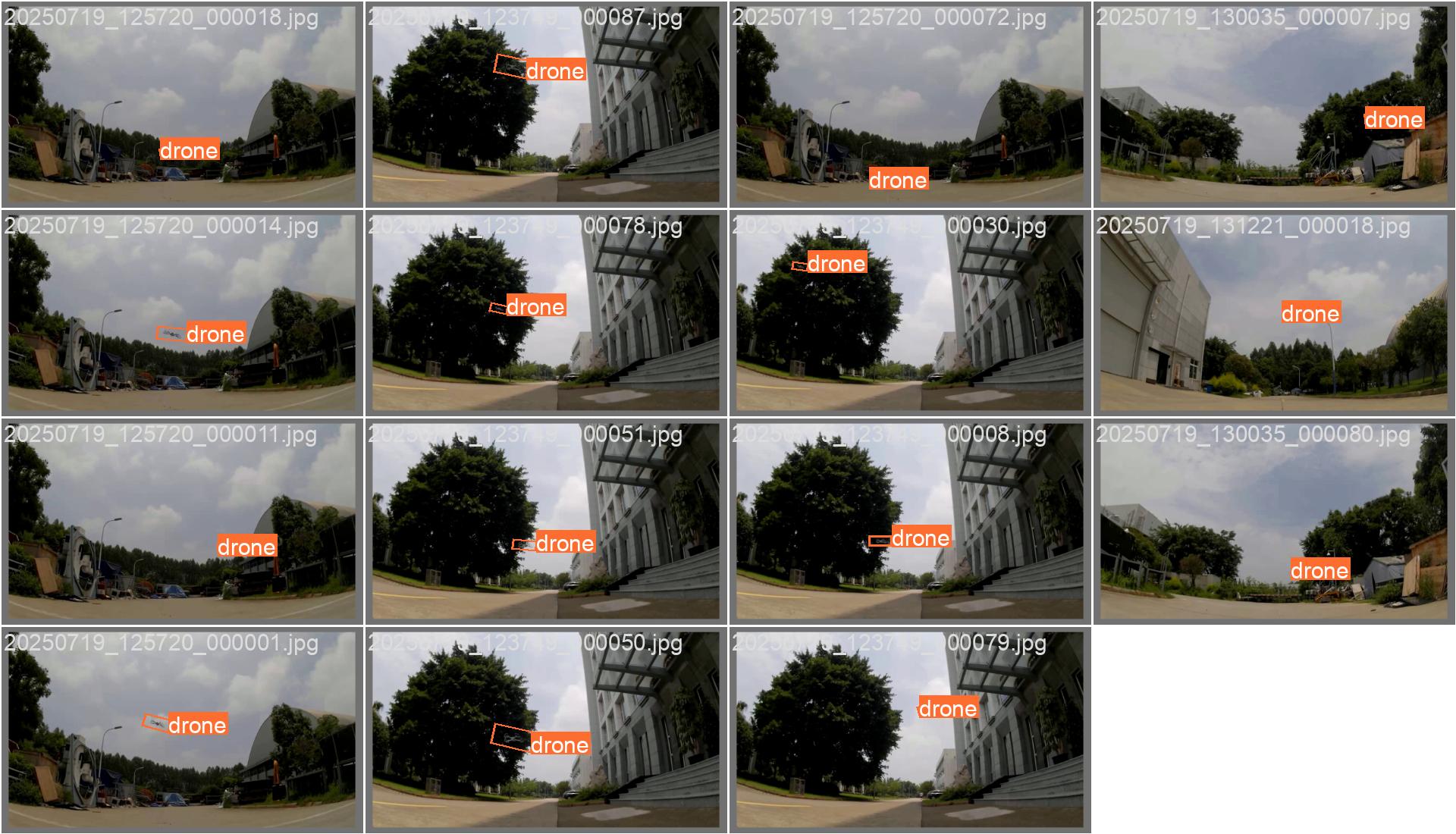}\\
        {\small (a) Validation labels.}
    \end{minipage}
    \hfill
    \begin{minipage}[t]{0.48\linewidth}
        \centering
        \includegraphics[width=\linewidth]{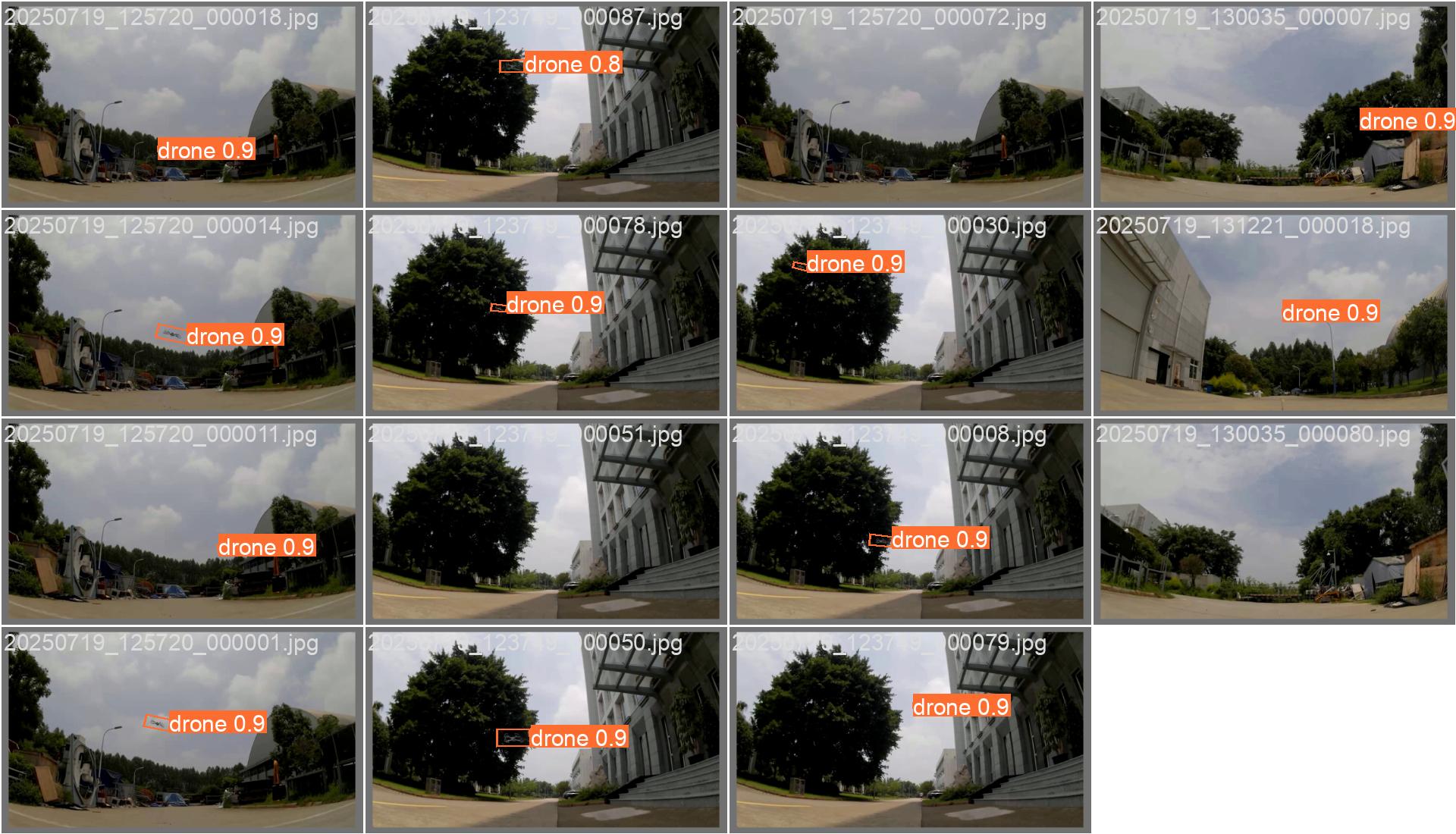}\\
        {\small (b) Model predictions.}
    \end{minipage}
    \caption{Qualitative examples of OBB labels and YOLO-OBB predictions.}
    \label{fig:val_vis}
\end{figure}

\begin{figure}[htbp]
    \centering
    \includegraphics[width=0.65\linewidth]{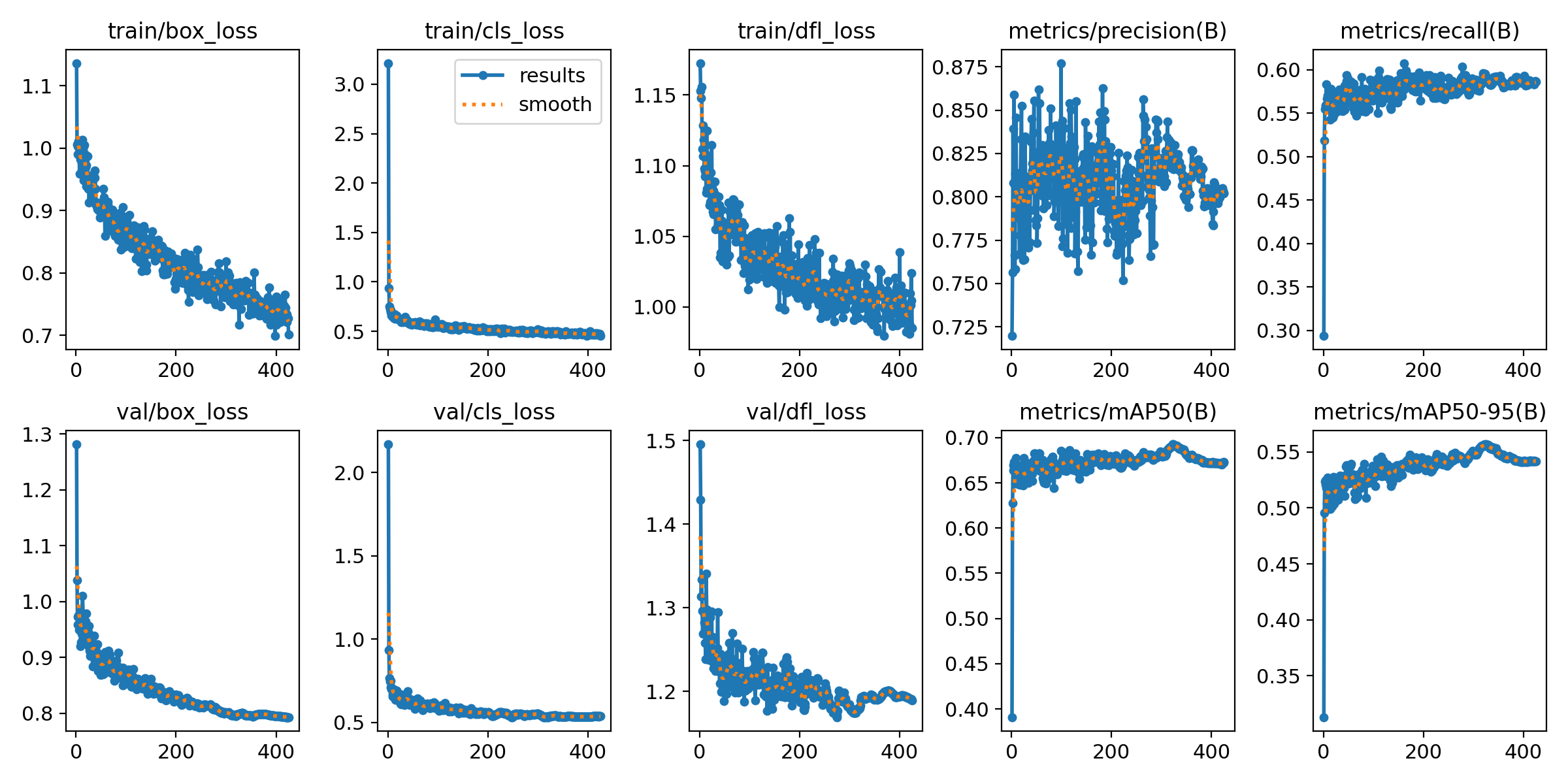}
    \caption{YOLO-OBB training curves.}
    \label{fig:train_curves}
\end{figure}

The detector is trained with an input resolution of $640\times640$.
The training curves in Fig.~\ref{fig:train_curves} show stable convergence.
The validation result reaches $\mathrm{mAP@0.5}=0.692$, and the UAV class reaches $\mathrm{AP}_{50}=0.973$.
This indicates that the weak-prior labels are sufficiently consistent for learning stable UAV OBB direction and scale.

\begin{figure}[htbp]
    \centering
    \includegraphics[width=0.92\linewidth]{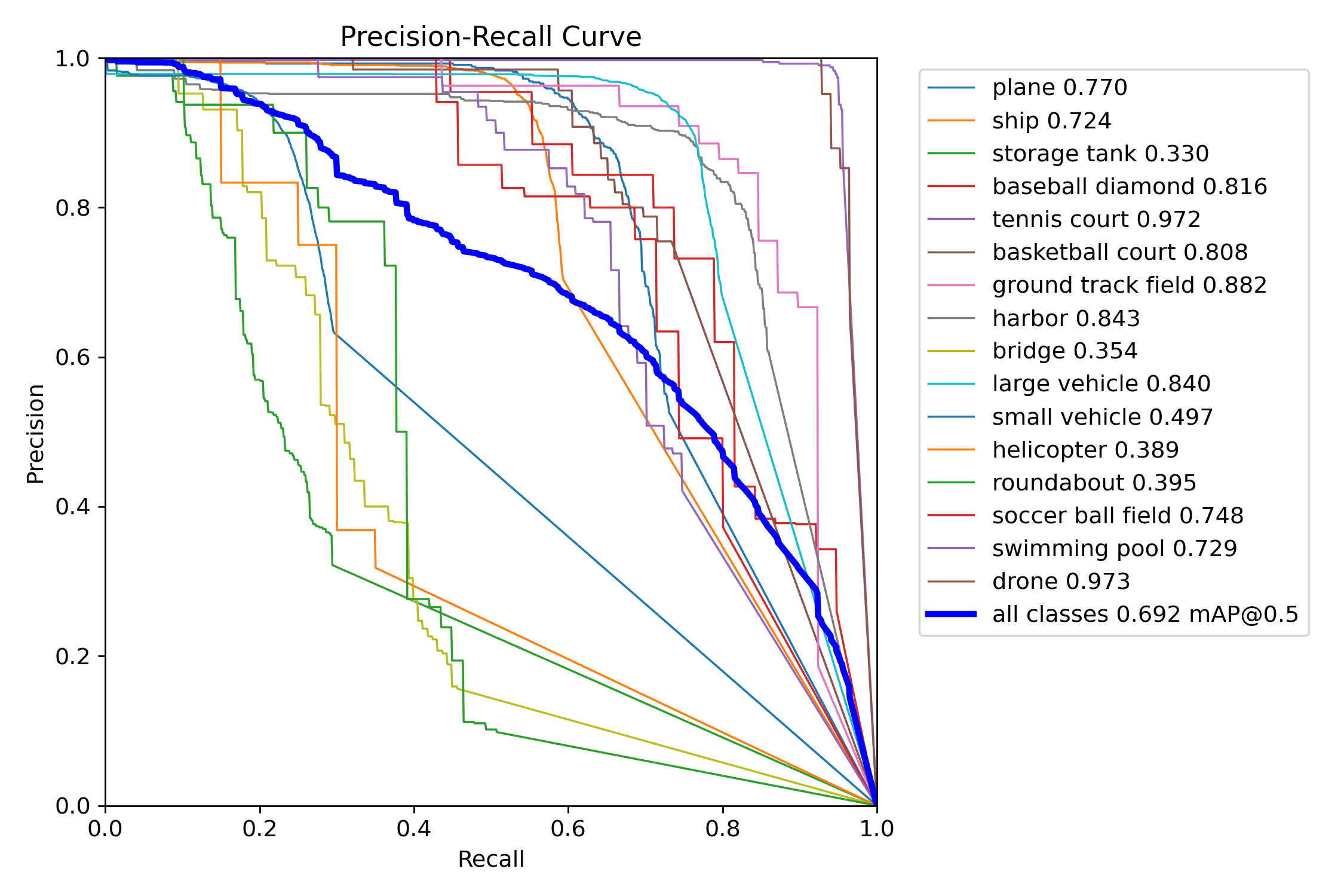}
    \caption{Validation precision--recall curves and per-class $\mathrm{AP}_{50}$.}
    \label{fig:pr_curve}
\end{figure}

\begin{figure}[htbp]
    \centering
    \includegraphics[width=0.92\linewidth]{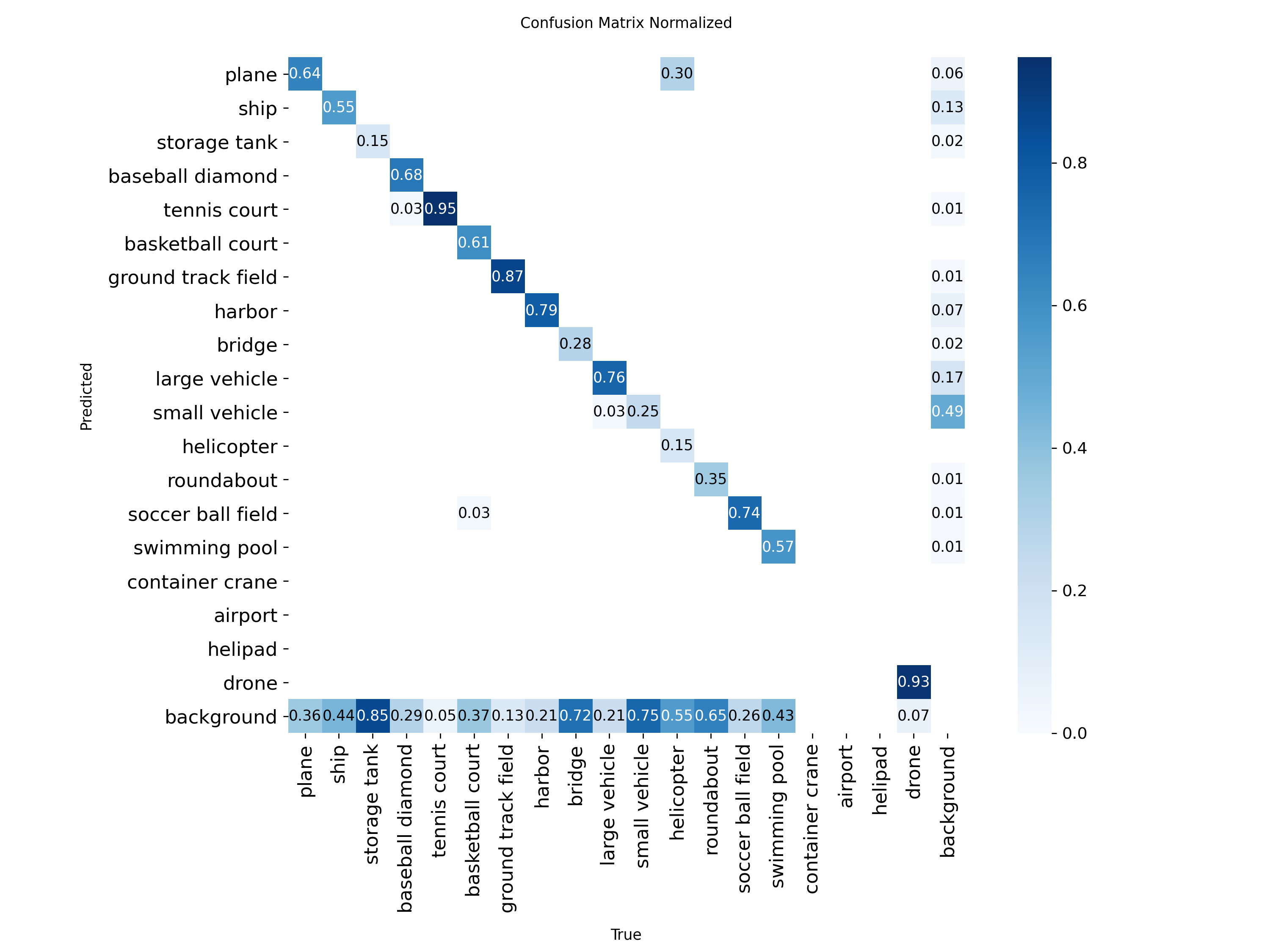}
    \caption{Normalized confusion matrix on the validation set.}
    \label{fig:confusion}
\end{figure}

\subsection{Simulation Setup}

The MATLAB simulation evaluates whether tilt observations improve short-horizon prediction.
The trajectory generator produces a nonlinear three-dimensional motion.
Each camera outputs
\begin{equation}
    \mathbf{z}^{(c)}_k
    =
    \begin{bmatrix}
        u^{(c)}_k &
        v^{(c)}_k &
        \rho^{(c)}_k &
        \eta^{(c)}_{{\rm roll},k} &
        \eta^{(c)}_{{\rm pitch},k}
    \end{bmatrix}^{\T}.
\end{equation}

Dropouts and false detections are generated by independent camera-level Markov chains.
For each camera, $T_i=1$ means that the true target is detected, and $F_i=1$ means that a false detection is present.
The three-camera detection state is
\begin{equation}
    \mathbf{s}_k
    =
    \begin{bmatrix}
        T_{1,k} & F_{1,k} &
        T_{2,k} & F_{2,k} &
        T_{3,k} & F_{3,k}
    \end{bmatrix}^{\T}.
\end{equation}
The steady-state true detection probability is $0.98$, the mean target-loss duration is $30$ frames, the steady-state false-detection probability is $0.02$, and the mean false-detection duration is $30$ frames.

Two settings are compared:
\begin{itemize}[leftmargin=*, itemsep=0.2em, topsep=0.2em]
    \item \textbf{Baseline:} image position and range are used; roll/pitch observations are disabled.
    \item \textbf{Roll/Pitch:} image position, range, and image-domain roll/pitch are used.
\end{itemize}
The prediction horizon is $\tau=0.5~\mathrm{s}$.
The prediction error is
\begin{equation}
    e(t_k)
    =
    \left\|
    \mathbf{p}_{\rm pred}(t_k)
    -
    \mathbf{p}_{\rm true}(t_k+\tau)
    \right\|_2 .
    \label{eq:pred_error}
\end{equation}

\subsection{Simulation Results}

Table~\ref{tab:sim_gain} summarizes the prediction errors.
Adding image-domain roll/pitch observations decreases the cumulative error from $6611.580~\mathrm{m\cdot s}$ to $2594.837~\mathrm{m\cdot s}$.
The RMSE decreases from $1.991~\mathrm{m}$ to $0.821~\mathrm{m}$, and the maximum error decreases from $6.556~\mathrm{m}$ to $3.553~\mathrm{m}$.

\begin{table}[htbp]
    \centering
    \caption{Prediction error reduction from image-domain roll/pitch observations in simulation.}
    \label{tab:sim_gain}
    \footnotesize
    \setlength{\tabcolsep}{8pt}
    \renewcommand{\arraystretch}{1.2}
    \begin{tabular}{lccc}
        \toprule
        Metric & Baseline & Roll/Pitch & Reduction \\
        \midrule
        Cumulative error $E_{\rm cum}$ (m$\cdot$s) & 6611.580 & 2594.837 & 60.75\%\\
        Mean error $\bar e$ (m)                    & 1.838    & 0.721    & 60.75\%\\
        RMSE (m)                                   & 1.991    & 0.821    & 58.73\%\\
        Maximum error $e_{\max}$ (m)               & 6.556    & 3.553    & 45.81\%\\
        \bottomrule
    \end{tabular}
\end{table}

\begin{figure}[htbp]
    \centering
    \includegraphics[width=0.92\linewidth]{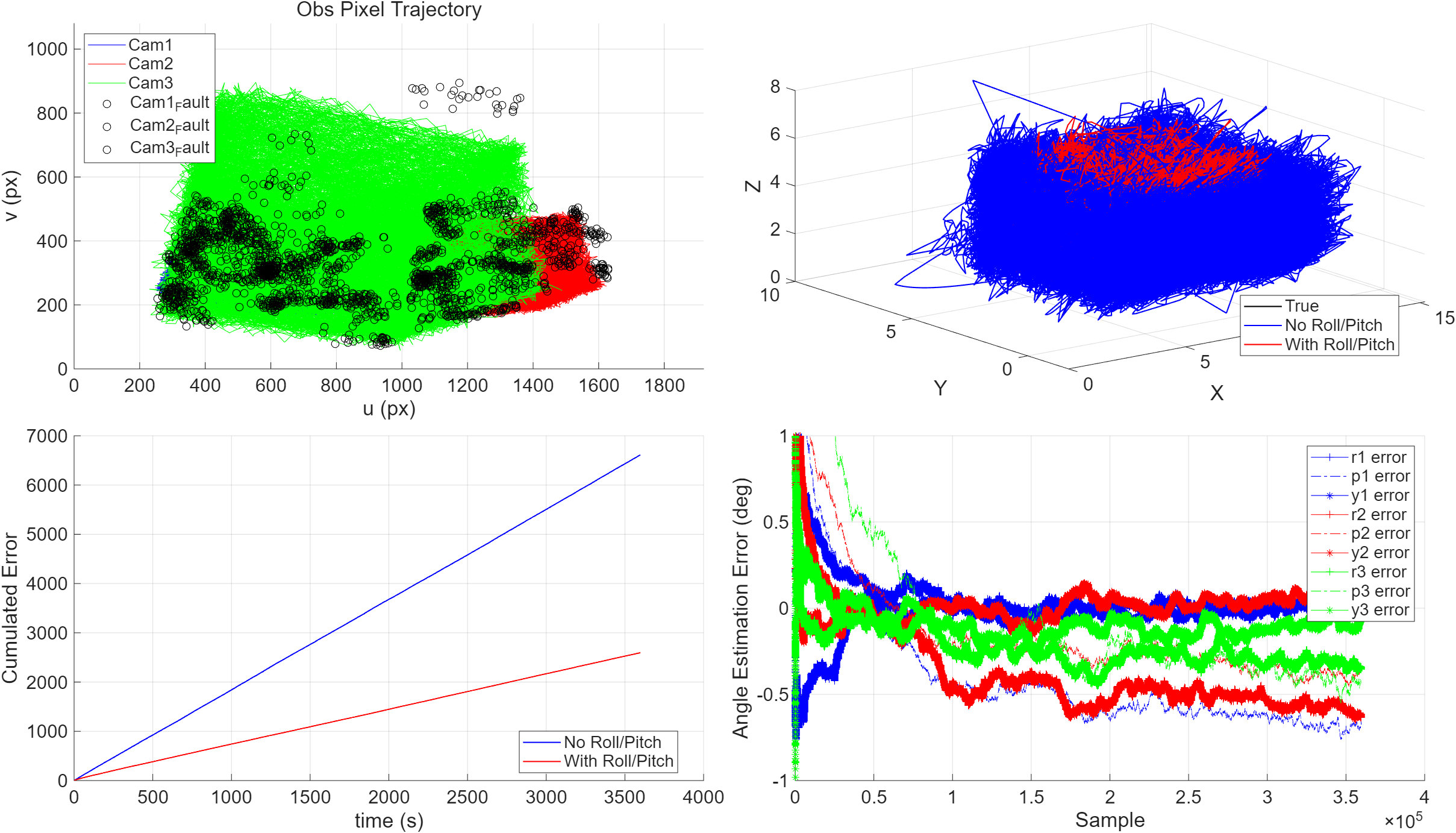}
    \caption{Overall results of the three-camera asynchronous simulation.
    Top-left: observed pixel trajectories with false detections.
    Top-right: 3D trajectory comparison.
    Bottom-left: cumulative prediction error.
    Bottom-right: camera Euler-angle estimation error.}
    \label{fig:sim_overall}
\end{figure}

\begin{figure}[htbp]
    \centering
    \includegraphics[width=0.92\linewidth]{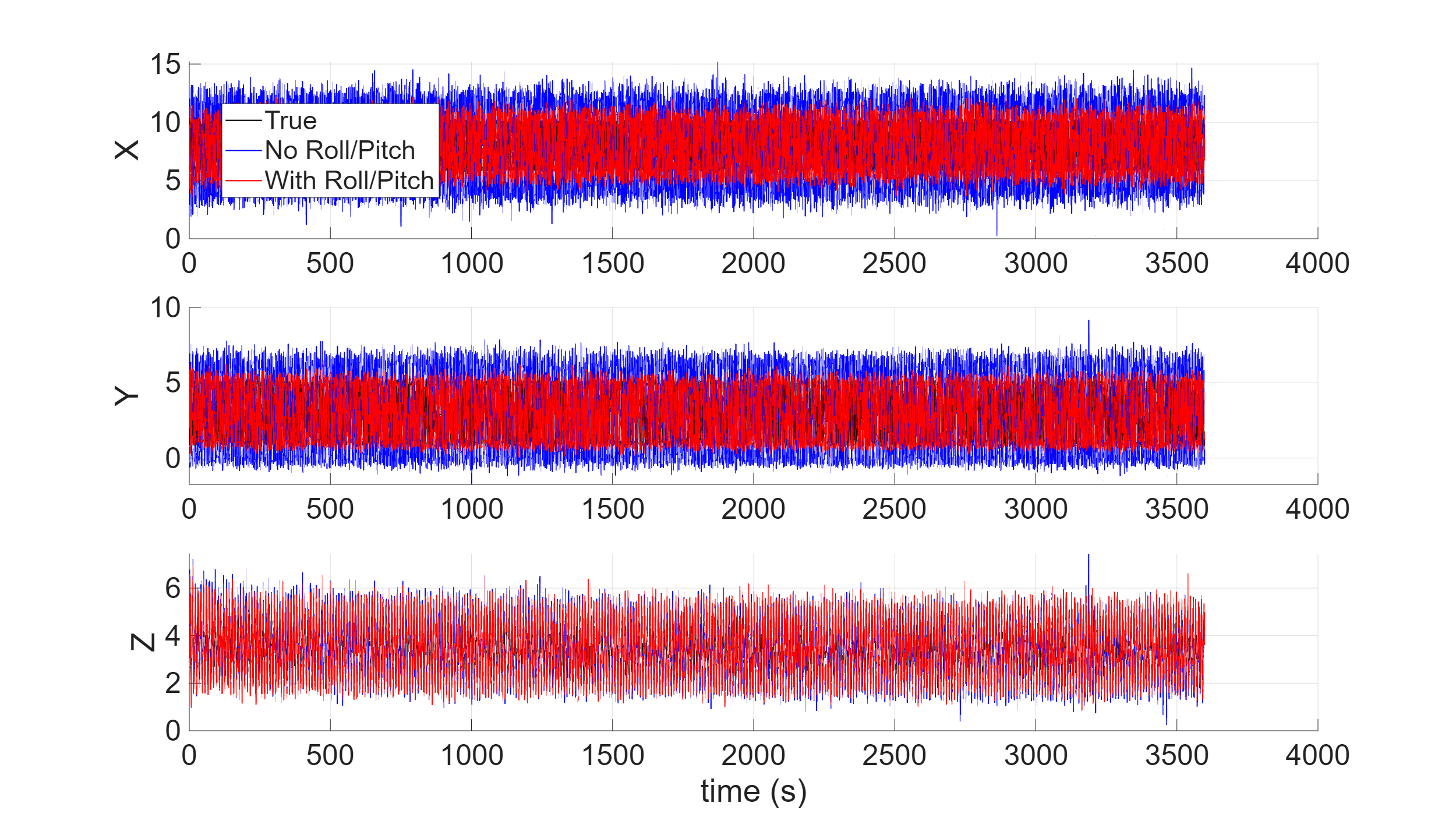}
    \caption{Time series comparison of 3D position.
    Black: true trajectory.
    Blue: baseline without roll/pitch.
    Red: filter with roll/pitch observations.}
    \label{fig:sim_xyz}
\end{figure}

\begin{figure}[htbp]
    \centering
    \includegraphics[width=0.92\linewidth]{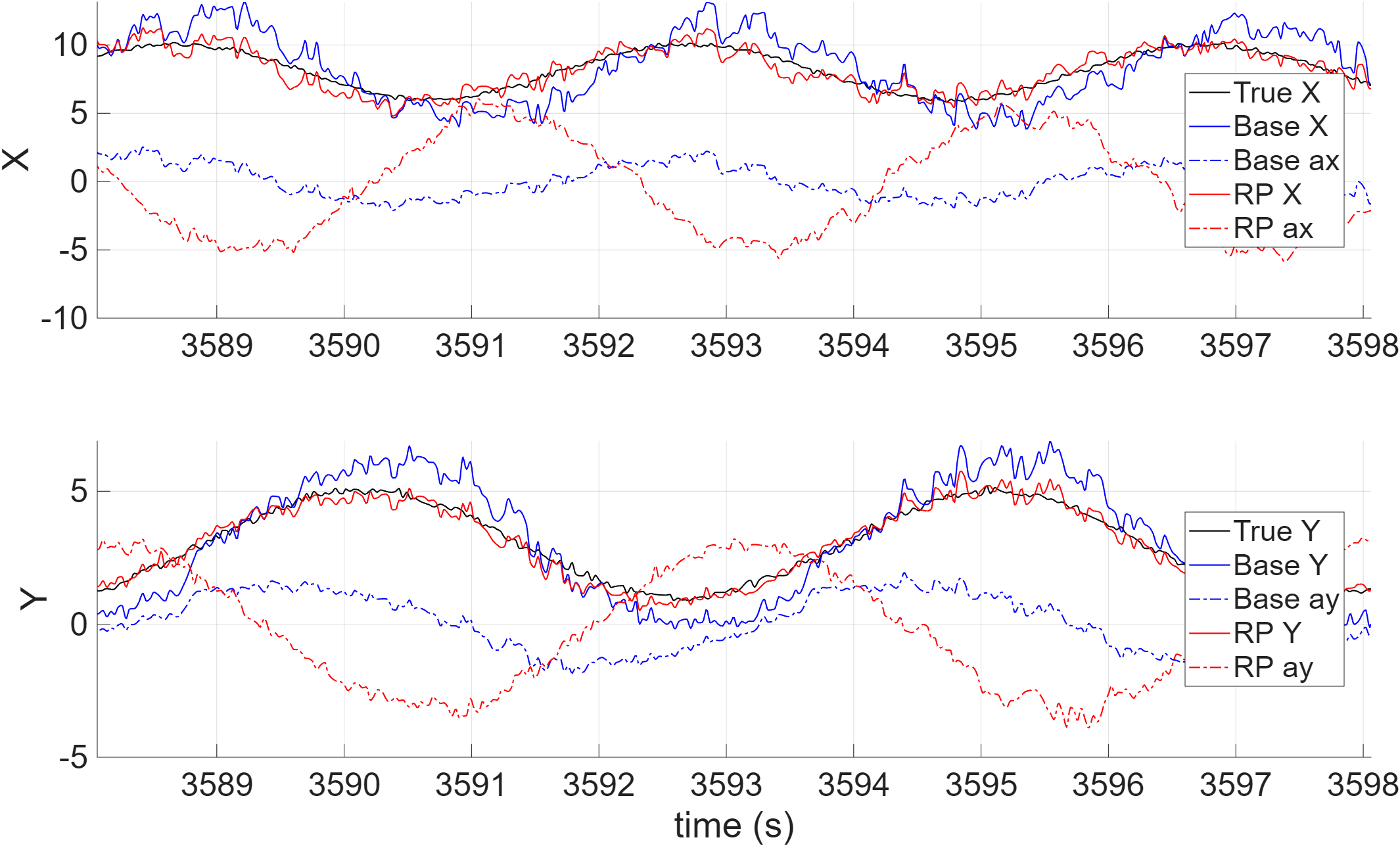}
    \caption{Position--acceleration comparison in the last $10~\mathrm{s}$ simulation window.}
    \label{fig:sim_last10}
\end{figure}

The simulation verifies the expected mechanism.
Image-domain roll/pitch directly constrains the acceleration-related part of the state, reducing prediction lag during maneuvering.

\subsection{Real Distributed Multi-Camera Experiment}

The real experiment uses one mobile gimbal camera mounted on a quadruped robot and two fixed ground cameras.
The mobile camera provides a reconfigurable viewpoint, while the fixed cameras provide stable redundant observations.
Each incoming observation is assigned to a fixed camera slot and fused asynchronously.

The real experiment does not use independent high-precision 3D ground truth.
Therefore, a self-consistency metric is used.
At time $t$, the state estimated at $t-\tau$ is predicted to $t$, and the predicted observation is compared with the current observation:
\begin{equation}
    e(t)
    =
    \sqrt{
    \frac{1}{|\Omega(t)|}
    \sum_{j\in\Omega(t)}
    r_j^2(t)
    },
    \label{eq:self_consistency}
\end{equation}
where
\begin{equation}
    r_j(t)
    =
    \begin{cases}
        \wrap\left(\hat{z}_j(t|t-\tau)-z_j(t)\right), & z_j \text{ is an angle},\\[0.3em]
        \hat{z}_j(t|t-\tau)-z_j(t), & z_j \text{ is non-angular}.
    \end{cases}
\end{equation}
The cumulative self-consistency error is
\begin{equation}
    E=\int e(t)\,dt.
\end{equation}

Table~\ref{tab:real_gain} shows the result.
Enabling roll/pitch observations reduces the cumulative self-consistency error from $203.4518$ to $166.6241$, corresponding to an $18.10\%$ reduction.
The improvement is smaller than in simulation because the real system contains detector jitter, residual extrinsic error, timing mismatch, and camera-dependent observation quality.
Nevertheless, the result shows that image-domain tilt remains useful under practical asynchronous multi-camera conditions.

\begin{table}[htbp]
    \centering
    \caption{Real distributed experiment: cumulative self-consistency prediction error.}
    \label{tab:real_gain}
    \begin{tabular}{lcc}
        \toprule
        Method & Cumulative error $E$ & Relative reduction \\
        \midrule
        Baseline without Roll/Pitch & 203.4518 & -- \\
        Roll/Pitch enabled          & 166.6241 & 18.10\% \\
        \bottomrule
    \end{tabular}
\end{table}

\begin{figure}[htbp]
    \centering
    \includegraphics[width=\linewidth]{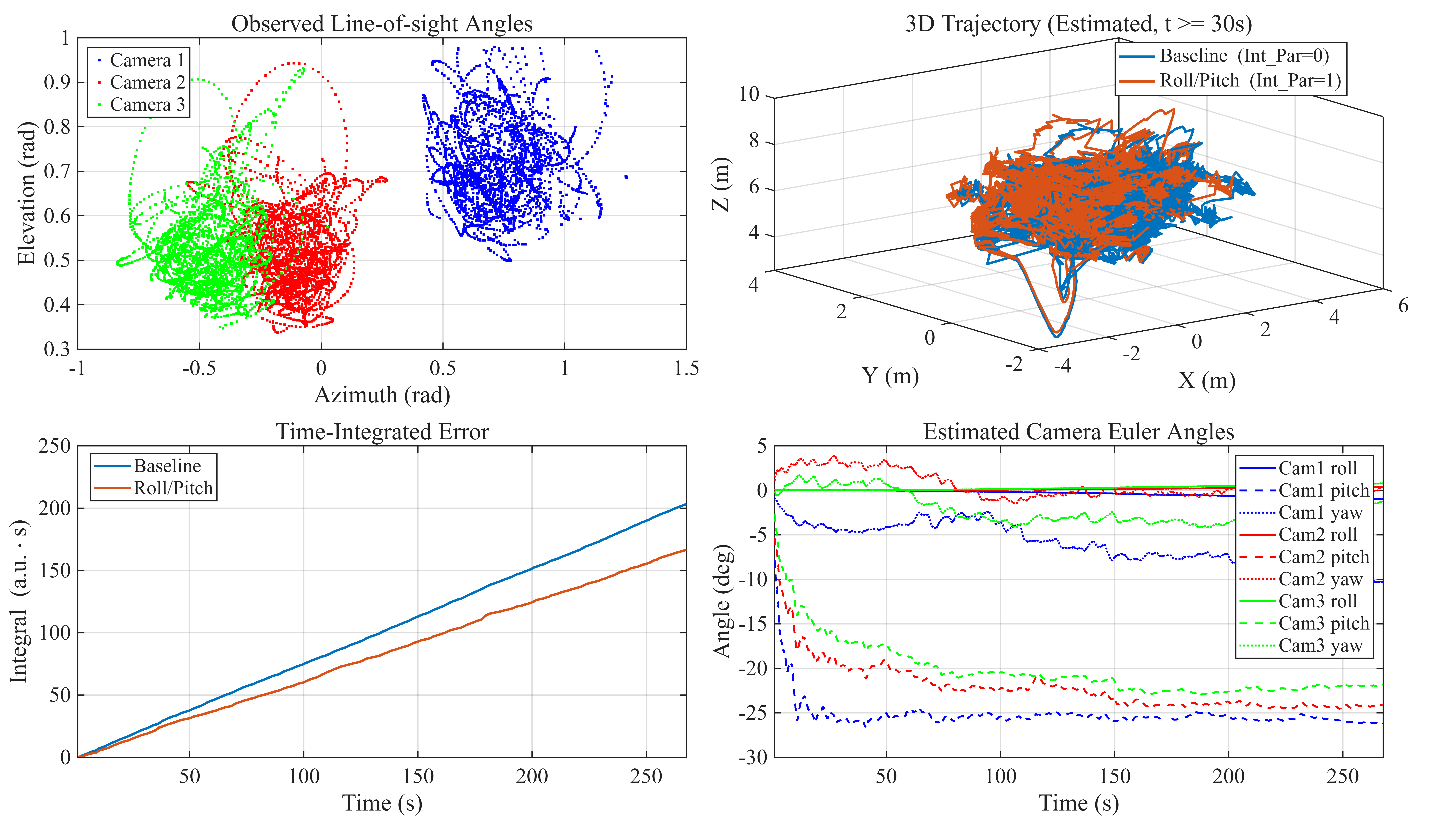}
    \caption{Overview of distributed fusion with a mobile gimbal camera and two fixed ground cameras.
    The figure includes line-of-sight observation distribution, 3D trajectory, cumulative error comparison, and estimated camera Euler-angle error states.}
    \label{fig:real_overview}
\end{figure}

\subsection{Robustness Analysis}

The main practical error sources are false detections, dropouts, aggressive maneuvers, and camera-pose inconsistency.
False detections are handled by the hard Mahalanobis gate.
Moderate residuals after dropouts are handled by covariance widening.
Aggressive maneuvers are handled by the tilt-to-acceleration pseudo-observation.
Camera-pose inconsistency is handled by the augmented camera attitude states.

\begin{figure}[htbp]
    \centering
    \includegraphics[width=.92\linewidth]{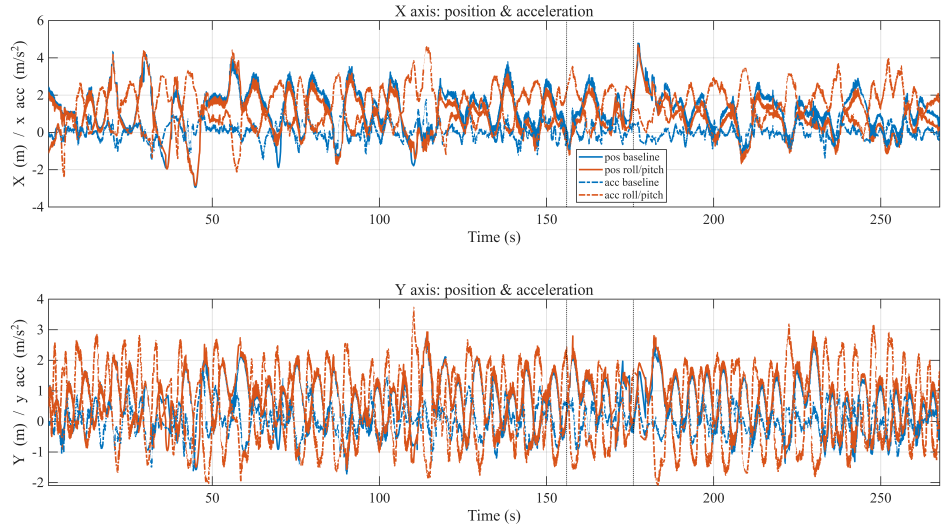}
    \caption{Distributed fusion position and acceleration comparison on $X/Y$ axes over the full sequence.
    Solid lines denote position and dash-dot lines denote acceleration.
    Blue denotes the baseline and red denotes the roll/pitch-enhanced filter.}
    \label{fig:real_posacc_full}
\end{figure}

\begin{figure}[htbp]
    \centering
    \includegraphics[width=.92\linewidth]{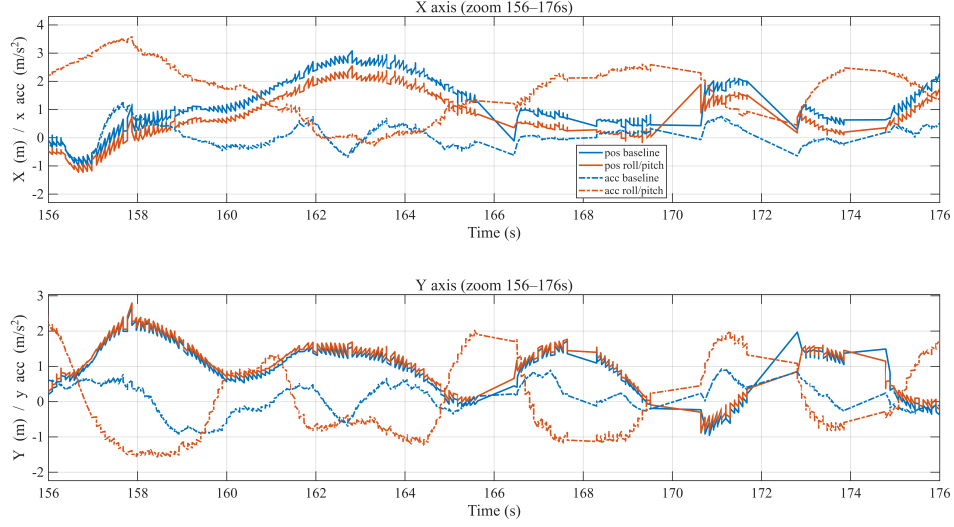}
    \caption{Zoomed aggressive maneuver segment from $156$ to $176~\mathrm{s}$.
    With roll/pitch observations, acceleration responds faster and oscillates less, improving short-horizon prediction consistency.}
    \label{fig:real_posacc_zoom}
\end{figure}

The zoomed maneuver segment illustrates the role of image-domain tilt.
Without tilt cues, acceleration is mainly inferred from position and range updates, so it tends to lag.
With roll/pitch cues, the filter receives an additional maneuver-related constraint and produces a more responsive acceleration estimate.

\section{Code Availability}

The Python implementation of the weak-prior auto-labeling pipeline and YOLO-OBB training workflow is publicly available at
\begin{center}
    \codeurl
\end{center}
The repository contains motion point extraction, target trajectory tracing, weak observed-angle and YOLO-OBB dataset generation, and YOLO-OBB training scripts.

\section{Conclusion}
\label{sec:conclusion}

This paper presented an image-domain tilt constrained distributed fusion framework for maneuvering UAV tracking.
The method uses synchronized video, gimbal IMU, and UAV IMU data to generate weak OBB and image-domain roll/pitch labels, trains a YOLO-OBB detector for online tilt extraction, and introduces the extracted tilt as an acceleration-related pseudo-observation in a nonlinear recursive estimator.
For distributed tracking, asynchronous observations from one mobile gimbal camera and two fixed ground cameras are fused with augmented camera attitude states.

The results show that image-domain tilt is a useful low-level maneuver cue.
In simulation, it reduces prediction RMSE by 58.73\% and cumulative prediction error by 60.75\%.
In the real multi-camera experiment, it reduces the self-consistency cumulative error by 18.10\%.
The smaller real-world gain is expected because the real system contains detector jitter, residual extrinsic errors, timing mismatch, and camera-dependent observation quality.
Future work will focus on improving tilt recognition stability, estimating time offsets explicitly, and jointly adapting detector confidence, tilt uncertainty, and distributed fusion consistency.

\bibliographystyle{IEEEtran}
\bibliography{drone}

@article{AQ2025,
  title = {Adaptive {{Kalman Filters Based}} on {{Elliptically Contoured Distributions}} for {{Heavy-Tailed}} and {{Nonstationary Measurement Noises}}},
  author = {Qi, Bin and Zhang, Songyuan and Chen, Weihan and Fu, Yili and Ren, Bingyin},
  year = 2025,
  journal = {IEEE Transactions on Instrumentation and Measurement},
  volume = {74},
  pages = {1--15},
  issn = {1557-9662},
  doi = {10.1109/TIM.2025.3625324},
  urldate = {2026-03-12}
}

@article{AT2024,
  title = {Adaptive {{Weighted Ridge Regression Estimator}} for {{Time-Varying Sensitivity Identification}}},
  author = {Tang, Zhiyuan and Liu, Youbo and Liu, Tingjian and Xu, Xiao and Liu, Junyong},
  year = 2024,
  month = jan,
  journal = {IEEE Transactions on Power Systems},
  volume = {39},
  number = {1},
  pages = {2377--2380},
  issn = {1558-0679},
  doi = {10.1109/TPWRS.2023.3329692},
  urldate = {2024-01-25}
}

@article{CA2009,
  title = {Cubature Kalman Filters},
  author = {Arasaratnam, Ienkaran and Haykin, Simon},
  year = 2009,
  month = jun,
  journal = {IEEE Transactions on Automatic Control},
  volume = {54},
  number = {6},
  pages = {1254--1269},
  issn = {1558-2523},
  doi = {10.1109/TAC.2009.2019800}
}

@article{CM2020,
  title = {Combination of {{IMM Algorithm}} and {{ASTRWCKF}} for {{Maneuvering Target Tracking}}},
  author = {Ma, Jian and Guo, Xiaoting},
  year = 2020,
  journal = {IEEE Access},
  volume = {8},
  pages = {143095--143103},
  issn = {2169-3536},
  doi = {10.1109/ACCESS.2020.3013561},
  urldate = {2024-01-22}
}

@inproceedings{DB2019,
  title = {Deep {{Convolutional Autoencoder}} for {{Estimation}} of {{Nonstationary Noise}} in {{Images}}},
  booktitle = {2019 8th {{European Workshop}} on {{Visual Information Processing}} ({{EUVIP}})},
  author = {Bahncmiri, Sheyda Ghanbaralizadeh and Ponomarenko, Mykola and Egiazarian, Karen},
  year = 2019,
  month = oct,
  pages = {238--243},
  issn = {2471-8963},
  doi = {10.1109/EUVIP47703.2019.8946273},
  urldate = {2026-03-12}
}

@article{DD2025,
  title = {Drone Detection in Airport Environments: {{A}} Literature Review},
  shorttitle = {Drone Detection in Airport Environments},
  author = {{de Macedo}, Sanderson Oliveira and Caetano, Mauro and {da Costa}, Ronaldo Martins},
  year = 2025,
  month = dec,
  journal = {Array},
  volume = {28},
  pages = {100511},
  issn = {2590-0056},
  doi = {10.1016/j.array.2025.100511},
  urldate = {2026-03-02}
}

@inproceedings{DK2022,
  title = {Detection and {{Recognition}} of {{Drones}} Using {{Deep Convolution Neural Networks}}},
  booktitle = {2022 {{IEEE}} 6th {{Conference}} on {{Information}} and {{Communication Technology}} ({{CICT}})},
  author = {Karthikeya Nalam, Venkata Sai and Amar Koushik Tanniru, Venkata Sai and Posani, Anjaneyulu and Suneetha, Manne},
  year = 2022,
  month = nov,
  pages = {1--5},
  doi = {10.1109/CICT56698.2022.9997921},
  urldate = {2026-03-02}
}

@article{DM2025,
  title = {Deep Learning Based Image Classification for Embedded Devices: {{A}} Systematic Review},
  shorttitle = {Deep Learning Based Image Classification for Embedded Devices},
  author = {Moreira, Larissa Ferreira Rodrigues and Moreira, Rodrigo and Travencolo, Bruno Augusto Nassif and Backes, Andre Ricardo},
  year = 2025,
  month = mar,
  journal = {Neurocomputing},
  volume = {623},
  pages = {129402},
  publisher = {Elsevier},
  address = {Amsterdam},
  issn = {0925-2312, 1872-8286},
  doi = {10.1016/j.neucom.2025.129402},
  langid = {english}
}

@article{DT1999,
  title = {The {{Discovery}} of {{Ceres}}: {{How Gauss Became Famous}}},
  shorttitle = {The {{Discovery}} of {{Ceres}}},
  author = {Teets, Donald and Whitehead, Karen},
  year = 1999,
  month = apr,
  journal = {Mathematics Magazine},
  volume = {72},
  number = {2},
  pages = {83--93},
  publisher = {Taylor \& Francis},
  issn = {0025-570X},
  doi = {10.1080/0025570X.1999.11996710},
  urldate = {2023-07-25}
}

@book{EB2001,
  title = {Estimation with Applications to Tracking and Navigation: Theory Algorithms and Software},
  author = {{Bar-Shalom}, Yaakov and Li, X Rong and Kirubarajan, Thiagalingam},
  year = 2001,
  publisher = {John Wiley \& Sons},
  isbn = {0-471-41655-X}
}

@inproceedings{ED2025,
  title = {Experimental {{Analysis}} of {{Fine-Tuned Drone Detection YOLO Models}}},
  booktitle = {2025 1st {{International Conference}} on {{Secure IoT}}, {{Assured}} and {{Trusted Computing}} ({{SATC}})},
  author = {Dogan, Sara Acikkol and Walatkiewicz, Jonathan and Tout, Samir and Darwish, Omar and Spantidi, Ourania},
  year = 2025,
  month = feb,
  pages = {1--5},
  doi = {10.1109/SATC65530.2025.11137294},
  urldate = {2026-03-02}
}

@article{ES2023,
  title = {An {{Ensemble-Based IoT-Enabled Drones Detection Scheme}} for a {{Safe Community}}},
  author = {Singh, Jaskaran and Sharma, Keshav and Wazid, Mohammad and Das, Ashok Kumar and Vasilakos, Athanasios V.},
  year = 2023,
  journal = {IEEE Open Journal of the Communications Society},
  volume = {4},
  pages = {1946--1956},
  issn = {2644-125X},
  doi = {10.1109/OJCOMS.2023.3310003},
  urldate = {2026-03-02}
}

@inproceedings{ET2010,
  title = {Exploration of Adaptive Filters for Target Tracking in the Presence of Model Uncertainty},
  booktitle = {2010 {{Sixth International Conference}} on {{Intelligent Sensors}}, {{Sensor Networks}} and {{Information Processing}}},
  author = {Truong, Tracy Q.S.},
  year = 2010,
  month = dec,
  pages = {1--6},
  publisher = {IEEE},
  doi = {10.1109/ISSNIP.2010.5706770}
}

@article{FR2017,
  title = {Faster {{R-CNN}}: {{Towards Real-Time Object Detection}} with {{Region Proposal Networks}}},
  shorttitle = {Faster {{R-CNN}}},
  author = {Ren, Shaoqing and He, Kaiming and Girshick, Ross and Sun, Jian},
  year = 2017,
  month = jun,
  journal = {IEEE Transactions on Pattern Analysis and Machine Intelligence},
  volume = {39},
  number = {6},
  pages = {1137--1149},
  issn = {1939-3539},
  doi = {10.1109/TPAMI.2016.2577031},
  urldate = {2026-03-02}
}

@article{FS2001,
  title = {A Framework for State-Space Estimation with Uncertain Models},
  author = {Sayed, A. H.},
  year = 2001,
  month = jul,
  journal = {IEEE TRANSACTIONS ON AUTOMATIC CONTROL},
  volume = {46},
  number = {7},
  pages = {998--1013},
  publisher = {Ieee-Inst Electrical Electronics Engineers Inc},
  address = {Piscataway},
  issn = {0018-9286, 1558-2523},
  doi = {10.1109/9.935054},
  urldate = {2023-07-23},
  langid = {english}
}

@article{FZ1981,
  title = {Feedback and Optimal Sensitivity: {{Model}} Reference Transformations, Multiplicative Seminorms, and Approximate Inverses},
  shorttitle = {Feedback and Optimal Sensitivity},
  author = {Zames, G.},
  year = 1981,
  month = apr,
  journal = {IEEE Transactions on Automatic Control},
  volume = {26},
  number = {2},
  pages = {301--320},
  issn = {1558-2523},
  doi = {10.1109/TAC.1981.1102603}
}

@misc{G,
  title = {Guide to Authors \textbar{} {{Nature Communications}}},
  urldate = {2024-12-10},
  url = {https://www.nature.com/ncomms/submit/guide-to-authors},
  note = {[EB/OL]}
}

@article{IB1988,
  title = {The {{Interacting Multiple Model Algorithm}} for {{Systems}} with {{Markovian Switching Coefficients}}},
  author = {Blom, Hap and Barshalom, Y.},
  year = 1988,
  month = aug,
  journal = {IEEE TRANSACTIONS ON AUTOMATIC CONTROL},
  volume = {33},
  number = {8},
  pages = {780--783},
  publisher = {Ieee-Inst Electrical Electronics Engineers Inc},
  address = {New York},
  issn = {0018-9286},
  doi = {10.1109/9.1299},
  urldate = {2023-07-23},
  langid = {english}
}

@article{LH2024,
  title = {Light-{{YOLOv5}}: {{A Lightweight Drone Detector}} for {{Resource-Constrained Cameras}}},
  shorttitle = {Light-{{YOLOv5}}},
  author = {Han, Jin and Cao, Ran and Brighente, Alessandro and Conti, Mauro},
  year = 2024,
  month = mar,
  journal = {IEEE Internet of Things Journal},
  volume = {11},
  number = {6},
  pages = {11046--11057},
  issn = {2327-4662},
  doi = {10.1109/JIOT.2023.3329221},
  urldate = {2026-03-02}
}

@incollection{LW1964,
  title = {The {{Linear Predictor}} for a {{Single Time Series}}},
  booktitle = {Extrapolation, {{Interpolation}}, and {{Smoothing}} of {{Stationary Time Series}}: {{With Engineering Applications}}},
  author = {Wiener, Norbert},
  year = 1964,
  pages = {56--80},
  publisher = {MIT Press},
  urldate = {2023-07-23},
  isbn = {978-0-262-25719-0}
}

@inproceedings{MB2018,
  title = {{{MIT Cheetah}} 3: {{Design}} and {{Control}} of a {{Robust}}, {{Dynamic Quadruped Robot}}},
  shorttitle = {{{MIT Cheetah}} 3},
  booktitle = {2018 {{IEEE}}/{{RSJ International Conference}} on {{Intelligent Robots}} and {{Systems}} ({{IROS}})},
  author = {Bledt, Gerardo and Powell, Matthew J. and Katz, Benjamin and Di Carlo, Jared and Wensing, Patrick M. and Kim, Sangbae},
  year = 2018,
  month = oct,
  pages = {2245--2252},
  issn = {2153-0866},
  doi = {10.1109/IROS.2018.8593885},
  urldate = {2026-03-05}
}

@inproceedings{MV2020,
  title = {{{MPC-based Controller}} with {{Terrain Insight}} for {{Dynamic Legged Locomotion}}},
  booktitle = {2020 {{IEEE International Conference}} on {{Robotics}} and {{Automation}} ({{ICRA}})},
  author = {Villarreal, Octavio and Barasuol, Victor and Wensing, Patrick M. and Caldwell, Darwin G. and Semini, Claudio},
  year = 2020,
  month = may,
  pages = {2436--2442},
  issn = {2577-087X},
  doi = {10.1109/ICRA40945.2020.9197312},
  urldate = {2024-06-02}
}

@article{NF2010,
  title = {New Interacting Multiple Model Algorithms for the Tracking of the Manoeuvring Target},
  author = {Fu, X. and Jia, Y. and Du, J. and Yu, F.},
  year = 2010,
  month = oct,
  journal = {IET CONTROL THEORY AND APPLICATIONS},
  volume = {4},
  number = {10},
  pages = {2184--2194},
  publisher = {Inst Engineering Technology-Iet},
  address = {Hertford},
  issn = {1751-8644},
  doi = {10.1049/iet-cta.2009.0583},
  urldate = {2023-09-01},
  langid = {english}
}

@article{NG1993,
  title = {Novel-{{Approach}} to {{Nonlinear Non-Gaussian Bayesian State Estimation}}},
  author = {Gordon, Nj and Salmond, Dj and Smith, Afm},
  year = 1993,
  month = apr,
  journal = {IEE PROCEEDINGS-F RADAR AND SIGNAL PROCESSING},
  volume = {140},
  number = {2},
  pages = {107--113},
  publisher = {Inst Engineering Technology-Iet},
  address = {Hertford},
  issn = {0956-375X},
  doi = {10.1049/ip-f-2.1993.0015},
  urldate = {2023-07-25},
  langid = {english}
}

@inproceedings{NJ1997,
  title = {A New Extension of the Kalman Filter to Nonlinear Systems},
  booktitle = {Signal {{Processing}}, {{Sensor Fusion}}, and {{Target Recognition Vi}}},
  author = {Julier, Simon J. and Uhlmann, Jeffrey K.},
  year = 1997,
  volume = {3068},
  pages = {182--193},
  publisher = {Spie - Int Soc Optical Engineering},
  address = {Bellingham},
  doi = {10.1117/12.280797},
  urldate = {2023-07-23},
  isbn = {978-0-8194-2483-9},
  langid = {english}
}

@article{NK1960,
  title = {A New Approach to Linear Filtering and Prediction Problems},
  author = {Kalman, R. E.},
  year = 1960,
  month = mar,
  journal = {Journal of Basic Engineering},
  volume = {82},
  number = {1},
  pages = {35--45},
  issn = {0021-9223},
  doi = {10.1115/1.3662552},
  urldate = {2023-07-23}
}

@inproceedings{OG2024,
  title = {Optimizing {{Real-Time Image Processing}} in {{Augmented Reality}} with {{Low-Latency Edge AI}}},
  booktitle = {2024 2nd {{International Conference}} on {{Signal Processing}}, {{Communication}}, {{Power}} and {{Embedded System}} ({{SCOPES}})},
  author = {{Gokila Deepa G} and {Gomathi S} and {Aadhitya S} and {Sujitha R} and {Sundarrajan M} and Choudhry, Mani Deepak},
  year = 2024,
  month = dec,
  pages = {1--6},
  doi = {10.1109/SCOPES64467.2024.10990829},
  urldate = {2026-03-12}
}

@inproceedings{PA2023,
  title = {Pre-Trained {{Deep Learning Networks}} for {{Advanced Visible Imagery Drone Detection}} and {{Recognition}}},
  booktitle = {2023 {{IEEE}} 15th {{International Conference}} on {{Computational Intelligence}} and {{Communication Networks}} ({{CICN}})},
  author = {{Al dawasari}, Hassan J. and Bilal, Muhammad and Moinuddin, Muhammad and Arshad, Kamran and Assaleh, Khaled},
  year = 2023,
  month = dec,
  pages = {316--320},
  issn = {2472-7555},
  doi = {10.1109/CICN59264.2023.10402291},
  urldate = {2026-03-02}
}

@incollection{PF2001,
  title = {Particle {{Filters}} for {{Mobile Robot Localization}}},
  booktitle = {Sequential {{Monte Carlo Methods}} in {{Practice}}},
  author = {Fox, Dieter and Thrun, Sebastian and Burgard, Wolfram and Dellaert, Frank},
  editor = {Doucet, Arnaud and {de Freitas}, Nando and Gordon, Neil},
  year = 2001,
  series = {Statistics for {{Engineering}} and {{Information Science}}},
  pages = {401--428},
  publisher = {Springer},
  address = {New York, NY},
  doi = {10.1007/978-1-4757-3437-9_19},
  urldate = {2023-07-25},
  isbn = {978-1-4757-3437-9},
  langid = {english}
}

@misc{PG2022,
  title = {Perceptive {{Locomotion}} through {{Nonlinear Model Predictive Control}}},
  author = {Grandia, Ruben and Jenelten, Fabian and Yang, Shaohui and Farshidian, Farbod and Hutter, Marco},
  year = 2022,
  month = aug,
  number = {arXiv:2208.08373},
  eprint = {2208.08373},
  primaryclass = {cs},
  publisher = {arXiv},
  doi = {10.48550/arXiv.2208.08373},
  urldate = {2024-06-02},
  archiveprefix = {arXiv}
}

@inproceedings{PJ2024,
  title = {Performance {{Comparison}} of {{YOLO Algorithms}} in {{Drone Detection}}},
  booktitle = {2024 {{IEEE International Conference}} on {{Smart Power Control}} and {{Renewable Energy}} ({{ICSPCRE}})},
  author = {Jain, Rakshit and Shrivastav, Sujal and Kakde, Sumit and Raut, Roshani},
  year = 2024,
  month = jul,
  pages = {1--6},
  doi = {10.1109/ICSPCRE62303.2024.10675042},
  urldate = {2026-03-02}
}

@article{RG2003,
  title = {Robust Adaptive Tracking for Time-Varying Uncertain Nonlinear Systems with Unknown Control Coefficients},
  author = {Ge, S. S. and Wang, J.},
  year = 2003,
  month = aug,
  journal = {IEEE TRANSACTIONS ON AUTOMATIC CONTROL},
  volume = {48},
  number = {8},
  pages = {1463--1469},
  publisher = {Ieee-Inst Electrical Electronics Engineers Inc},
  address = {Piscataway},
  issn = {0018-9286},
  doi = {10.1109/TAC.2003.815049},
  urldate = {2023-07-25},
  langid = {english}
}

@incollection{RH1992,
  title = {Robust {{Estimation}} of a {{Location Parameter}}},
  booktitle = {Breakthroughs in {{Statistics}}},
  author = {Huber, Peter J.},
  editor = {Kotz, Samuel and Johnson, Norman L.},
  year = 1992,
  pages = {492--518},
  publisher = {Springer New York},
  address = {New York, NY},
  doi = {10.1007/978-1-4612-4380-9_35},
  urldate = {2023-07-24},
  isbn = {978-0-387-94039-7 978-1-4612-4380-9}
}

@article{RJ2018,
  title = {{Robust Gaussian-sum ensemble Kalman filter and its application in bearings-only tracking}},
  author = {Jiang, Hao-nan and Cai, Yuan-li},
  year = 2018,
  month = feb,
  journal = {Kongzhi Lilun yu Yingyong = Control Theory \& Applications},
  volume = {35},
  number = {2},
  publisher = {South China University of Technology},
  issn = {1000-8152},
  doi = {10.7641/CTA.2017.70116},
  urldate = {2023-07-24},
  copyright = {Copyright South China University of Technology Feb 2018},
  langid = {chinese}
}

@inproceedings{RL2022b,
  title = {Research on the Anti-{{UAV}} Distributed System for Airports : {{YOLOv5-based}} Auto-Targeting Device},
  shorttitle = {Research on the Anti-{{UAV}} Distributed System for Airports},
  booktitle = {2022 3rd {{International Conference}} on {{Computer Vision}}, {{Image}} and {{Deep Learning}} \& {{International Conference}} on {{Computer Engineering}} and {{Applications}} ({{CVIDL}} \& {{ICCEA}})},
  author = {Liu, Ruixi and Xiao, Yuxin and Li, Zhidong and Cao, Hanlin},
  year = 2022,
  month = may,
  pages = {864--867},
  doi = {10.1109/CVIDLICCEA56201.2022.9824842},
  urldate = {2026-03-02}
}

@article{RQ2025,
  title = {{Review of Event Camera-Based Target Detection and Tracking Algorithms}},
  author = {Qiu, Jiayu and Zhang, Yasheng and Fang, Yuqiang and Li, Pengju and Zheng, Kaiyuan},
  year = 2025,
  month = feb,
  journal = {Laser \& Optoelectronics Progress},
  volume = {62},
  number = {4},
  pages = {0400004},
  publisher = {Shanghai Inst Optics \& Fine Mechanics, Chinese Acad Science},
  address = {Shanghai},
  issn = {1006-4125},
  doi = {10.3788/LOP241073},
  langid = {chinese}
}

@inproceedings{RS2025,
  title = {Real-{{Time UAV Detection Using}} an {{Enhanced YOLO}} v8 {{Model}}},
  booktitle = {2025 42nd {{National Radio Science Conference}} ({{NRSC}})},
  author = {Serageldin, Ahmed F. and Elsayed, Hussein A. and {Abdel-Hamid}, Lamiaa},
  year = 2025,
  month = may,
  volume = {1},
  pages = {193--201},
  issn = {2837-018X},
  doi = {10.1109/NRSC65659.2025.11018566},
  urldate = {2026-03-02}
}

@article{RW2015,
  title = {{Robust cubature Kalman filter target tracking algorithm based on genernalized M-estiamtion}},
  author = {Wu, Hao and Chen, Shu-Xin and Yang, Bin-Feng and Chen, Kun},
  year = 2015,
  month = nov,
  journal = {ACTA PHYSICA SINICA},
  volume = {64},
  number = {21},
  pages = {218401},
  publisher = {Chinese Physical Soc},
  address = {Beijing},
  issn = {1000-3290},
  doi = {10.7498/aps.64.218401},
  urldate = {2023-07-24},
  langid = {chinese}
}

@article{SC2026,
  title = {Stochastic {{Event-Triggered Robust Tracking Algorithm Under Nonstationary Heavy-Tailed Noise}} and {{Packet Dropouts}}},
  author = {Chen, Yu and Cai, Yuanli and Deng, Yifan and Liu, Jiaqi},
  year = 2026,
  journal = {IEEE Transactions on Automation Science and Engineering},
  volume = {23},
  pages = {4428--4441},
  issn = {1558-3783},
  doi = {10.1109/TASE.2026.3661491},
  urldate = {2026-03-12}
}

@inproceedings{ST2025,
  title = {Small {{Object Detection}} and {{Classification Using YOLO}} on {{Multi-Dataset Drone Imagery}}},
  booktitle = {2025 {{IEEE Pune Section International Conference}} ({{PuneCon}})},
  author = {Taware, Vaishnavi and Bhandare, Mohit and Bhirud, Sunil and Sawant, Suraj and Joshi, Amit},
  year = 2025,
  month = dec,
  pages = {1--6},
  issn = {2831-5022},
  doi = {10.1109/PuneCon67554.2025.11378548},
  urldate = {2026-03-02}
}

@article{UJ2004,
  title = {Unscented Filtering and Nonlinear Estimation},
  author = {Julier, S.J. and Uhlmann, J.K.},
  year = 2004,
  month = mar,
  journal = {Proceedings of the IEEE},
  volume = {92},
  number = {3},
  pages = {401--422},
  issn = {1558-2256},
  doi = {10.1109/JPROC.2003.823141}
}

@inproceedings{UV2000,
  title = {The {{Unscented Particle Filter}}},
  booktitle = {Advances in {{Neural Information Processing Systems}}},
  author = {{van der Merwe}, Rudolph and Doucet, Arnaud and {de Freitas}, Nando and Wan, Eric},
  year = 2000,
  volume = {13},
  publisher = {MIT Press},
  urldate = {2023-07-25}
}

@inproceedings{UW2000,
  title = {The Unscented Kalman Filter for Nonlinear Estimation},
  booktitle = {Proceedings of the {{IEEE}} 2000 {{Adaptive Systems}} for {{Signal Processing}}, {{Communications}}, and {{Control Symposium}} ({{Cat}}. {{No}}.{{00EX373}})},
  author = {Wan, E.A. and Van Der Merwe, R.},
  year = 2000,
  month = oct,
  pages = {153--158},
  doi = {10.1109/ASSPCC.2000.882463}
}

\end{document}